\documentclass[a4paper,11pt]{article}
\usepackage{jcappub} % for details on the use of the package, please
\usepackage[utf8]{inputenc}
\usepackage[T1]{fontenc} % if needed

\usepackage{multicol}
\usepackage{placeins}

\usepackage{subcaption}

\newcommand{\chris}[1]{{\bfseries\textcolor{red}{#1}}}

\newcommand{\eg}{e.g.}
\newcommand{\ie}{i.e.}

\newcommand{\mpbh}{m_{\rm PBH}}
\newcommand{\fpbh}{f_{\rm PBH}}

\newcommand{\li}{{\rm Li}}

\newcommand{\comment}[1]{}

\newcommand{\lambdabar}{{\mkern0.75mu\mathchar '26\mkern -9.75mu\lambda}}

\usepackage{bm}

\title{Revisiting constraints on asteroid-mass primordial black holes as dark matter candidates}

\author[a,b]{Paulo Montero-Camacho,}
\author[c]{Xiao Fang,}
\author[a,b]{Gabriel Vasquez,}
\author[a,b]{Makana Silva,}
\author[a,b,d]{and Christopher M. Hirata}
\affiliation[a]{Center for Cosmology and AstroParticle Physics, The Ohio State University, 191 West Woodruff Avenue, Columbus, OH 43210, USA}
\affiliation[b]{Department of Physics, The Ohio State University, 191 West Woodruff Avenue, Columbus, OH 43210, USA}
\affiliation[c]{Department of Astronomy and Steward Observatory, University of Arizona, 933 N Cherry Ave, Tucson, AZ 85719, USA}
\affiliation[d]{Department of Astronomy, The Ohio State University, 140 West 18th Avenue, Columbus, OH 43210, USA}
\emailAdd{monterocamacho.1@osu.edu}
\emailAdd{xfang@email.arizona.edu}
\emailAdd{silva.179@osu.edu}
\emailAdd{vasquez.119@osu.edu}
\emailAdd{hirata.10@osu.edu}

\abstract{As the only dark matter candidate that does not invoke a new particle that survives to the present day, primordial black holes (PBHs) have drawn increasing attention recently. Up to now, various observations have strongly constrained most of the mass range for PBHs, leaving only small windows where PBHs could make up a substantial fraction of the dark matter. Here we revisit the PBH constraints for the asteroid-mass window, \ie, the mass range $3.5\times 10^{-17}M_\odot < \mpbh < 4\times 10^{-12}M_\odot$. We revisit 3 categories of constraints. (1) For optical microlensing, we analyze the finite source size and diffractive effects and discuss the scaling relations between the event rate, $m_{\rm PBH}$ and the event duration. We argue that it will be difficult to push the existing optical microlensing constraints to much lower $\mpbh$. (2) For dynamical capture of PBHs in stars, we derive a general result on the capture rate based on phase space arguments. We argue that survival of stars does not constrain PBHs, but that disruption of stars by captured PBHs should occur and that the asteroid-mass PBH hypothesis could be constrained if we can work out the observational signature of this process. (3) For destruction of white dwarfs by PBHs that pass through the white dwarf without getting gravitationally captured, but which produce a shock that ignites carbon fusion, we perform a 1+1D hydrodynamic simulation to explore the post-shock temperature and relevant timescales, and again we find this constraint to be ineffective. In summary, we find that the asteroid-mass window, which was previously constrained due to femtolensing, WD survival, optical microlensing, and neutron star capture is no longer constrained. Hence, the asteroid-mass window remains open for PBHs to account for all the dark matter.}


\begin{document}
\maketitle
\flushbottom

\section{Introduction}
\label{sec:intro}

In the $\Lambda$CDM model, roughly $\Omega_{\rm DM} \simeq 0.26$ of the Universe's total energy density is made of dark matter (DM) \cite{2016A&A...594A..13P}, whose nature remains enigmatic even though evidence for its existence was first reported over 80 years ago \cite{1937ApJ....86..217Z}. Since then a wide range of astrophysical observations have pointed toward the existence of dark matter \cite{2005PhR...405..279B}, such as the galactic rotation curves \cite{1970ApJ...159..379R}, the baryon density constrained by the Big Bang Nucleosynthesis (see \cite{2009PhR...472....1I} for a recent review), the anisotropies in the Cosmic Microwave Background (CMB) and the inhomogeneities in the large-scale structure \cite{2003moco.book.....D,1985ApJ...292..371D, 2006Natur.440.1137S, 2016A&A...594A..13P}. 
In addition, gravitational lensing provides direct measurements of the mass distribution, hence becoming a powerful probe of DM, which is further developed into several techniques: the strong gravitational lensing by massive galaxy clusters \cite{1998ApJ...501..539T}, the weak gravitational lensing of galaxies by galaxies and large-scale structure \cite{2003ARA&A..41..645R, 2006PhR...429....1L,2014MNRAS.442.1326K,2014MNRAS.441.2725F,2017MNRAS.465.1454H,2017arXiv170801530D,2017arXiv170801538T}, and CMB lensing \cite{2016A&A...594A..13P,2014JCAP...04..014D,2012ApJ...756..142V}. Also see \cite{2017arXiv171206615B} for a recent review of gravitational probes of DM physics.

Since dark matter cannot be composed of any of the Standard Model (SM) particles, the dominant paradigm is that it is a new type of particle. Such a particle would need to be stable over the lifetime of the Universe, have sufficiently weak interactions with the Standard Model particles that it has not yet been discovered, and have a viable production mechanism in the early Universe. Considerable efforts have been devoted to looking for DM with direct and indirect detection methods using particle experiments \cite{2015PrPNP..85....1K, 2016JPhG...43a3001M, 2017arXiv171201391P, 2016JPhG...43d4001B}. The scenario that has received the greatest attention is the thermal WIMP scenario (see \cite{2017arXiv170307364A, 2017arXiv170706277R} for recent reviews), in which the dark matter is a massive particle that was in thermal equilibrium in the early Universe when the temperature was much higher than the WIMP mass. The comoving number density of WIMPs decreases exponentially as the temperature drops below the WIMP mass and eventually WIMPs are so diluted that they can not annihilate with each other efficiently, thus ``freezing out''. A symmetry (e.g., $R$-parity in supersymmetric models) protects the WIMP from decay into SM particles. However, there are many other Beyond the Standard Model candidates for dark matter, each with its own phenomenology and observational/experimental signatures.

Primordial black holes (PBHs) have been considered as an alternative scenario for DM for almost fifty years \cite{1971MNRAS.152...75H, 1974MNRAS.168..399C, 1975ApJ...201....1C, 1975Natur.253..251C}. PBHs would be formed in the early Universe by gravitational collapse sourced by an order unity perturbation that makes the surrounding region collapse into a black hole (see \cite{2005astro.ph.11743C} for a brief review of diverse PBH formation mechanisms; there are many possibilities, but a source of perturbations beyond the extrapolation of the inflationary power spectrum is needed). There are several reasons for interest in PBHs as DM candidates. First, they are solutions to general relativity, and thus are the only DM scenario that does not invoke a new elementary particle that survives to the present day. Second is the related fact that their properties are highly constrained: in general relativity, a PBH's properties are determined by its mass $\mpbh$ and spin parameter $a_{\star,\rm PBH}$, and there are no additional free parameters needed to determine its interactions with visible matter. (For most of the constraints on PBHs, it is $\mpbh$ rather than $a_{\star,\rm PBH}$ that matters.) Finally, the parameter space for PBHs is inherently bounded at both the high-mass and low-mass ends. An obvious maximum mass is set by the observed astrophysical objects that are made of dark matter. A bound on the minimum mass is that in order to survive to the present, PBHs must be massive enough to not have completely evaporated via Hawking radiation \cite{1974MNRAS.168..399C}, which requires $\mpbh>2.5\times 10^{-19}\,M_\odot$ \cite{1991PhRvD..44..376M}.
 
Recent discoveries of binary black hole mergers with $\sim 10\,M_\odot$ \cite{2016PhRvL.116f1102A, 2017PhRvL.118v1101A, 2017ApJ...851L..35A, 2017PhRvL.119n1101A} have posed challenges to stellar evolution theories, reviving the enthusiasm about PBHs as DM in that mass range \cite{2016PhRvL.116t1301B,2016PhRvL.117f1101S,2017PDU....15..142C,2016arXiv160404932E,2016PhRvD..94b3516R,2017PhRvD..96l3523A, 2017JCAP...09..037R, 2018PhRvL.121h1304A}. However, there are significant constraints on PBHs in various mass ranges. The CMB spectral distortions and anisotropies strongly constrain monochromatic masses of PBH due to the non-blackbody spectrum from the additional energy injection by accreting PBHs \cite{2017PhRvD..95d3534A,2017PhRvD..96h3524P,2018PhRvD..97d3525N,2019arXiv190106809A}. (For spinning black holes, there may also be a superradiance constraint \cite{2013PhRvD..88d1301P}.) The sizes and velocity dispersions of stellar clusters at the cores of various ultra-faint dwarf galaxies have been proposed to independently constrain the stellar-mass PBH fraction due to the dynamical heating \citep[\eg][]{2016ApJ...824L..31B,2017PhRvL.119d1102K}. Halo wide binaries can constrain the mass distribution of potential PBHs because they are susceptible to gravitational perturbations due to encounters \cite{2014ApJ...790..159M}. The impact of Poisson noise from PBHs in the small-scale Lyman-$\alpha$ forest power spectrum has been recently utilized to constrain massive PBHs ($m_{\rm PBH} > 60 M_{\odot}$)   \cite{2019arXiv190310509M}. Both the Poisson noise and accretion methods have promise for future -- and current -- 21 cm observations \cite{2013MNRAS.435.3001T, 2017JCAP...08..017G, 2018PhRvD..98b3503H}. 
Furthermore, microlensing searches \cite{1986ApJ...304....1P} have constrained PBH masses in the planetary-to-stellar-mass window by monitoring stars in the Large Magellanic Cloud (MACHO \cite{2000ApJ...542..281A,2001ApJ...550L.169A}, EROS-2 \cite{2007A&A...469..387T}) and the Galactic Bulge \cite{2019arXiv190107120N}. Lower masses have been constrained by monitoring stars in the Kepler field \cite{2014ApJ...786..158G} and stars in M31 observed by Subaru/HSC \cite{2019NatAs.tmp..238N}. It has also been shown in \cite{2018PhRvL.121n1101Z} that one can constrain PBH masses in the range $\geq 0.01 M_{\odot}$ from lensing of Type Ia SNe light curves due to PBHs. At the low masses, there are constraints from bursts of Hawking radiation from the final stages of the PBH evaporation \cite{1976ApJ...206....1P,2018ApJ...857...49A}, low-energy leptonic cosmic rays \cite{2019PhRvL.122d1104B}, and from the $\gamma$-ray background \cite{2010PhRvD..81j4019C}. Normally these constraints are plotted in the $(\mpbh,\fpbh)$ plane, where $\fpbh=\Omega_{\rm PBH}/\Omega_{\rm DM}$ is the fraction of the dark matter in PBHs. Note that all these constraints have been made assuming monochromatic mass functions, although considering extended mass functions do not relax those constraints \cite{2016PhRvD..94f3530G,2017PhRvD..95h3508K, 2017PhRvD..96b3514C, 2018JCAP...04..007L}. 

There is also an asteroid-mass window for PBH dark matter, where the mass is too large for constraints from Hawking radiation ($\gtrsim 5\times 10^{-17} M_\odot$) but too small for current optical microlensing surveys ($\lesssim 4\times 10^{-12} M_\odot$). In this work we will revisit the PBH constraints in this mass range and correct or strengthen them. The relevant constraints are from the femtolensing and picolensing of GRBs; microlensing of stars in M31 \cite{2019NatAs.tmp..238N}; the dynamical capture of PBHs by stars, including neutron stars (NS) and white dwarfs (WD) \cite{2013PhRvD..87b3507C,2013PhRvD..87l3524C,2014JCAP...06..026P,2014PhRvD..90h3507C}; and WD survival since passage of a PBH could ignite carbon fusion and destroy the WD \cite{2015PhRvD..92f3007G}. We note that recently asteroid-mass PBHs have been considered as a possible explanation of the positron cosmic ray excess \cite{2019PhLB..789..538T}, adding another motivation for constraining this mass window. 

Gammay ray bursts (GRBs) are interesting for asteroid-mass PBHs because of their cosmological distance (hence large lensing probability) and the short wavelength of electromagnetic radiation (so that the Einstein radius can exceed the Fresnel radius even for low-mass lenses). Femtolensing occurs when a GRB source is strongly lensed by an intervening PBH and the two light paths interfere with each other, leaving a signature in the GRB energy spectrum -- although the lensed images are not individually resolved \cite{1992ApJ...386L...5G}. Picolensing occurs when the magnification varies on $\sim$AU scales in the observer plane, and hence the fluence of gamma rays is different as seen at different interplanetary spacecraft \cite{1995ApJ...452L.111N}. Upper limits on PBH dark matter are obtained from non-observation of these phenomena. There have been reported constraints on PBHs in the $2.5\times 10^{-16}M_\odot<\mpbh<5\times 10^{-14}M_\odot$ range from femtolensing \cite{2012PhRvD..86d3001B}, however due to finite source size effects this range of PBH masses is now allowed \cite{2018JCAP...12..005K}. There is one picolensing constraint that includes BATSE + Ulysses \cite{1999ApJ...512L..13M}, although it used an older cosmological model and depends on an uncertain source redshift; it is not clear whether with modern parameters $\fpbh=1$ would be excluded for any value of $\mpbh$. Thus, at present, there is not an excluded range of $\mpbh$ from GRBs.

This paper is devoted to a thorough re-consideration of the other (non-GRB) constraints on the asteroid-mass PBH window. We place particular emphasis on optical microlensing toward M31 (where we outline the scaling laws that set the event rate and discuss the implications for future constraints); dynamical capture by stars (where we derive a general inequality for the PBH capture rate); and ignition of white dwarfs (where we use 1D hydrodynamics to follow WD material following passage of a PBH, instead of relying on order-of-magnitude arguments as in previous work). These constraints previously excluded PBHs in the asteroid-mass window as all of the dark matter.

In \S\ref{sec:micro}, we revisit optical microlensing events as a PBH constraint using optical diffractive and finite source size approximations that were flagged in Ref.~\cite{2019NatAs.tmp..238N} (see also Ref.~\cite{2019arXiv190506066S}). These considerations become important as one tries to utilize microlensing to constrain PBH masses below $10^{-10}M_\odot$, where the Einstein radii of the PBHs in the source plane are comparable to the stellar radii. We also analyze the microlensing event rate in the small mass limit for $\mpbh$ ($\mpbh < 10^{-10}M_{\odot}$) and find simple scaling relations with the $\mpbh$ and duration of an event in both the geometric and diffractive cases.

In \S\ref{sec:capture}, we discuss the issues with constraining PBHs with the dynamical capture of PBHs by stars. We derive a general result on the rate of PBH capture based on phase space arguments, and conclude that the survival of stars (in the sense of requiring the vast majority of stars to have survived a Hubble time) does not constrain PBHs. However, we find that {\em if} asteroid-mass PBHs are the dark matter, then capture and disruption of stars {\em must} happen, albeit at a rate 3--4 times smaller than that for supernovae. We speculate on the possible observational signatures of such events, which might lead to a future constraint.

In \S\ref{sec:explosion}, we examine whether the collision between a PBH and a WD will lead to thermonuclear explosions of the WD. Going beyond the calculations in \cite{2015PhRvD..92f3007G}, we carefully calculate the trajectory of a PBH passing through a WD and the involved micro-physics present during the passage. In particular, we perform a $1 + 1$D Lagrangian hydrodynamic simulation to track the thermodynamics of the surrounding WD materials and to find whether a shock forms. We then consider the conditions for a shock to start a runaway explosion based on the comparison of nuclear burning to conduction times \cite{1992ApJ...396..649T}, and discuss the uncertainties associated with hydrodynamic instabilities following PBH passage (which are very different from the considerations in Type Ia supernova ignition).

We discuss all of these results, and the outlook for asteroid-mass PBH dark matter, in \S\ref{sec:con}. 

\section{Optical microlensing}
\label{sec:micro}

Optical microlensing is the principal tool to constrain dark compact objects in the planetary-to-stellar mass range. Historically, searches aimed at constraining such objects as dark matter candidates have been carried out toward the Magellanic Clouds. In particular, EROS observations \cite{2007A&A...469..387T} ruled out the possibility that PBHs make up all of the dark matter if their mass is in the range from $7\times 10^{-8}$ to $15\,M_\odot$ (the lower range is constrained by the faster-cadence EROS-1 CCD survey). MACHO observations rule out the 0.3--30$\,M_\odot$ window \cite{2001ApJ...550L.169A} (though see recent discussion on the precise upper limit \cite{2017PhRvD..96d3020G}).

One natural question is whether optical microlensing can probe lower mass ranges. While the microlensing optical depth is independent of $\mpbh$, there are challenges at the lower masses because the finite sizes of the source stars become important, and even if the stars were pointlike the event durations become very short. One approach is to accept the finite source size effects, and aim for very high precision photometry so that small amplifications $A$ with $A-1\ll 1$ can be detected. This has been done with {\slshape Kepler} \cite{2014ApJ...786..158G}, resulting in a constraint that PBHs in the mass range from $2\times 10^{-9}$ to $10^{-7}$ $M_\odot$ cannot make up all of the dark matter. The other approach is to use high-cadence observations of sources that are much farther away, e.g., in M31.

The advent of wide-field cameras on large telescopes makes this second approach very promising, and indeed PBH dark matter has been constrained with the HSC observations of M31 \cite{2019NatAs.tmp..238N}. These observations set very strong upper bounds on PBH dark matter -- indeed, the reported constraints go down all the way to $4\times 10^{-12}\,M_\odot$, with the reach of the search at low $\mpbh$ set by the finite source size effects and diffraction effects.

In this section, we discuss the anticipated microlensing event rates at such small $\mpbh$, including finite source size and diffraction. Our main goal is to understand the relevant scaling relations in this regime (which is qualitatively different from other microlensing applications), and how this affects the detectability of PBHs, both in the HSC M31 survey and in possible future surveys using a similar technique.

\subsection{Computation of the microlensing event rate}
\label{ss:muRate}

From geometrical considerations, the microlensing event rate per star if all dark matter is in black holes of mass $\mpbh$ is
\begin{equation}
{\rm d}\Gamma = \frac{2\rho_{\rm DM}(D_{\rm OL})}{\mpbh}v_\perp P(v_\perp|D_{\rm OL}) \, {\rm d}D_{\rm OL} \, {\rm d}b \,{\rm d}v_\perp,
\end{equation}
where $v_\perp$ is the transverse velocity of the lens relative to the observer-source line, $D_{\rm OL}$ is the observer-lens distance, and $P(v_\perp|D_{\rm OL})$ is the conditional probability distribution for the transverse velocity. The impact parameter $b$ is measured in the lens plane. The factor of 2 arises since lenses can pass ``above'' or ``below'' the observer-source line. The formal limits of integration are for $0<D_{\rm OL}<D_{\rm OS}$, $b>0$, and $v_\perp>0$. In practice, the event rate will be finite because events with some maximum impact parameter can be detected. We thus write
\begin{equation}
{\rm d}\Gamma = \frac{2\rho_{\rm DM}(D_{\rm OL})}{\mpbh}v_\perp b_{\rm max}(v_\perp,D_{\rm OL}) P(v_\perp|D_{\rm OL}) \, {\rm d}D_{\rm OL} \, {\rm d}v_\perp.
\label{eq:Gamma-bv}
\end{equation}
Formally, we take $b_{\rm max}=0$ if an event would be undetectable regardless of impact parameter.

It is common to take a Maxwellian velocity distribution for the dark matter; if we do this, and take the standard assumption that there is a dispersion of $\sigma = v_c/\sqrt2$ per axis and a mean velocity of $\bar v_\perp$ (relative to the observer-source line of sight), then we find that
\begin{equation}
P(v_\perp) = \frac{v_\perp}{\sigma^2} e^{-(v_\perp^2 + \bar v_\perp^2)/2\sigma^2} I_0\left( \frac{\bar v_\perp v_\perp}{\sigma^2} \right),
\end{equation}
% 2D: P(v) = 1/(2 pi sigma^2) * e^{-v^2/2sigma^2}
where $I_0$ is the modified Bessel function of the first kind.
We take a standard circular velocity of $v_c = 230$ km s$^{-1}$. For the mean transverse velocity, we assume a non-rotating DM halo and consider the mean transverse velocity to arise from the rotation of the disk: then
\begin{equation}
\bar v_\perp = \frac{D_{\rm LS}}{D_{\rm OS}} v_c \sqrt{1-\cos^2{b}\sin^2{l}},
\end{equation}
where $l$ is the Galactic longitude and $b$ is the Galactic latitude. The distance ratio accounts for the fact that transverse velocity is measured in the lens plane.

We tried both using (a) the MW dark matter density model used in \S\ref{ss:runaway} (see Ref.~\cite{2017JCAP...02..007B}), and (b) the same model used in the HSC M31 analysis (the uncontracted NFW model from Ref.~\cite{2002ApJ...573..597K}). The differences are modest; typically model (b) leads to event rates that are $\sim$10\% lower, and in what follows we use model (b). The profile is cut off at the viral radius, $r_{\rm vir} = 258$ kpc. We also included the M31 dark matter halo\footnote{Again, following the HSC M31 analysis, we used the uncontracted $C_1$ model from Ref.~\cite{2002ApJ...573..597K}. We assume the source is 8 kpc from the center of M31 and in the plane of the sky.}, but found it made no difference for $\mpbh\lesssim 10^{-9} M_\odot$ and $A_{\rm min}=1.34$, since lenses in M31 have Einstein radii in the source plane that are smaller than the source stars.

To estimate the maximum impact parameter, we make a simple cut on the microlensing events: we require that the amplification $A$ (ratio of lensed to unlensed flux) exceed some threshold value $A_{\rm min}$ for at least a duration of time $t_{\rm min}$.\footnote{This gives the cumulative distribution in time, whereas the differential distribution ${\rm d}\Gamma/{\rm d}t$ is given in Ref.~\cite{2017JCAP...02..007B}; the two are of course related by the fundamental theorem of calculus.} The impact parameter will be related to (but not identical to) the Einstein radius in the lens plane,
\begin{equation}
R_{\rm E,L} = \sqrt{\frac{4\pi G \mpbh D_{\rm OL}D_{\rm LS}}{c^2D_{\rm OS}}}.
\end{equation}
In the geometric point source limit, the amplification is well known to be
\begin{equation}
A(x) = \frac{2+x^2}{x\sqrt{4+x^2}},
\label{eq:A-of-x}
\end{equation}
where $x = r_\perp/R_{\rm E,L}$ is the ratio of the transverse separation to the Einstein radius in the lens plane. In this case, we could say that given $A_{\rm min}$, there is some maximum value of $x$, $x_{\rm max}$, for which the amplification exceeds $A$. The maximum impact parameter for which the amplification exceeds $A_{\rm min}$ for a duration of at least $t_{\rm min}$ is then given by the Pythagorean theorem:
\begin{equation}
b_{\rm max} = \sqrt{\max \left( R_{\rm E,L}^2x_{\rm max}^2 - \frac{v_\perp^2t_{\rm min}^2}{4} , 0\right)}.
\label{eq:bmax}
\end{equation}

Finite source size effects are parameterized by the ratio of the source radius to the Einstein radius in the source plane, $r\equiv R_{\rm S}/R_{\rm E,S}$, where $R_{\rm E,S} = R_{\rm E,L} D_{\rm OS}/D_{\rm OL}$. For simplicity, we ignore limb darkening and take a uniform surface brightness disc for the source. Following the logic of Ref.~\cite{1994ApJ...430..505W}, we may replace Eq.~(\ref{eq:A-of-x}) with
\begin{equation}
A(x,r) = \frac1{\pi r^2} \oint \frac12 \left( |z_+|^2 - |z_-|^2 \right) \,{\rm d}\psi,
\label{eq:A-of-xr}
\end{equation}
where the integral is taken over the edge of the source. This is simply a ratio of areas, analogous to Eq.~(4) of Ref.~\cite{1994ApJ...430..505W}, but written in polar coordinates. The quantities in the integral can be written in turn as
\begin{equation}
|z_\pm| = \frac12|\zeta| \left| 1 \pm \frac4{|\zeta|^2} \right|,~~
|\zeta| = \sqrt{x^2 + r^2 + 2xr\cos\varphi}, ~~{\rm and}~~
\psi = \tan^{-1} \frac{r\sin\varphi}{x+r\cos\varphi}.
\end{equation}
Here $\phi$ is the position angle of a point $P$ the source limb relative to the source center $(r,0)$; $|\psi|$ is the position angle of $P$ relative to the PBH at $(0,0)$; $|\zeta|$ is the transverse distance from $P$ to the PBH; $|z_\pm|$ are the distances from $P$ to the PBH; and all of these intermediate quantities have been scaled in Einstein units in the relevant plane. The integral is performed parametrically taking the independent variable $\varphi$ to range from $0$ to $2\pi$ (technically $0$ to $\pi$ and then doubling), rather than using the elliptic function formulation of Ref.~\cite{1994ApJ...430..505W}. Equations~(\ref{eq:bmax}) and (\ref{eq:A-of-xr}) then suffice to obtain $b_{\rm max}$ in the presence of finite source size.

Some computed event rates are shown in Figure~\ref{fig:deltacomp}. We show the case of a point source ($R_{\rm S} = 0$) for comparison in the left column, and then the cases of a finite source size star in the center column. We note that at the distance of M31, main sequences of magnitude $r=23$, 24, 25, and 26 would have a radii of 3.8, 3.4, 2.5, and $2.0R_\odot$ respectively; these cases are shown in the different rows.\footnote{The absolute magnitude for $r=26$ at a distance of 770 kpc would be $M_r = 1.57$. We converted this to a radius using the main sequence table of Ref.~\cite{2013ApJS..208....9P}, with extended fields ({\tt http://www.pas.rochester.edu/\~{}emamajek/EEM\_dwarf\_UBVIJHK\_colors\_Teff.txt}), converted to $ugriz$ photometry using color transformations \cite{2005AJ....130..873J}.} Both the left and center columns ignore diffractive effects (which are treated in \S\ref{ss:diffraction}, and included in the right column).

\subsection{Diffractive effects}
\label{ss:diffraction}

We now consider diffractive effects in optical microlensing toward M31. These have also been discussed extensively by a recent paper \cite{2019arXiv190506066S}; our conclusions are broadly similar, in particular the finding that diffraction severely limits the ability for the M31 technique to probe very low masses.

We use the well-established result that for a distant observer approximation, the lensing magnification for a monochromatic point source can be expressed in terms of the $_1F_1$ confluent hypergeometric function with complex arguments \cite{1986ApJ...307...30D}. We may then obtain the lensing magnification for a finite broadband source by integrating over positions on the source and over wavenumbers. For the simple case of a top-hat bandpass, and a blackbody spectrum with no limb darkening, the resulting magnification is
\begin{equation}
 A  = \frac{\iiint dr\, d\phi\, dy \, 2\pi r y^{3} | _{1}F_{1}(iy,1,ix^{2}y)|^{2}(1-e^{-2\pi y})^{-1}(e^{y/y_{0}}-1)^{-1}}{\iiint dr\, d\phi\, dy \, r  y^{2}(e^{y/y_{0}}-1)^{-1}},
\end{equation}
where $A$ is averaged over positions and wavelengths, $r$ and $\phi$ are the radial and angular coordinates with respect to this coordinate system with the lens at the origin, $x$ is the position of an infinitesimal element of the source in the plane (a function of $r$ and $\phi$), $y$ is the dimensionless wavelength scale $y= 4\pi G\mpbh/(\lambda c^2)$, $y_0= 4\pi G\mpbh/(\lambda_0 c^2)$ and $\lambda_{0} = hc/kT_\star$ is the characteristic thermal wavelength of the star's photosphere (temperature $T_\star$). The wavelength limits are 5500--7000 \AA, corresponding to the $r2$ filter for the Hyper Suprime-Cam (HSC). We used a 6000 K blackbody spectrum; the mean value $\langle y\rangle$ only changes by 1.4\%\ even for a Rayleigh-Jeans spectrum ($T_\star=\infty$), so we did not re-compute the tables for each stellar temperature.

%We now consider $A_{\rm min}$. Physically, this is some minimum cut-off magnification based on the sensitivity of the telescope (i.e. HSC) being used to determine the effective impact parameter for the microlensing event rate.  It is expected that the effective impact parameter that depends on $\left< A_{\rm min} \right>$ will be different from the impact parameter derived from pure geometric optics, which implies that the region of observation will also change along with the event rate ($\Gamma$).

A subtlety is that in the diffractive case, $A$ is not a monotonically decreasing function of $x$: $A(x)$ often undergoes oscillations as $x\rightarrow \infty$ and $A\rightarrow 1$. Therefore, there are multiple solutions when we invert the function $A(x)$ to solve for $x_{\rm max}$ in terms of $A_{\rm min}$. Here we have chosen the largest value of $x_{\rm max}$: thus we are calculating the event rate $\Gamma$ for microlensing events where the time when $A$ first exceeds $A_{\rm min}$ to when $A$ last exceeds $A_{\rm min}$ is at least $t_{\rm min}$. Depending on the selection cuts in a particular survey, not all of these events will be detected; this calculation thus represents an optimistic bound on the reach of a microlensing survey.

In Fig.~\ref{fig:deltacomp}, the event rates with diffraction are plotted as a function of $t_{\rm min}$ in the right column. This can be compared to the event rates with only geometric considerations in the middle column of Fig.~\ref{fig:deltacomp}.

%\begin{figure}
%\centering
%\includegraphics[width=\textwidth]{rateplot.eps}
%\vspace{-15pt}
%\caption{\label{fig:event-rates}The cumulative microlensing event rates for $A>A_{\rm min}$, $t>t_{\rm min}$, and sources in M31 of the stated size. Lenses in both the Milky Way and M31 are considered. The upper panels are for point sources ($R=0$; the magnifications are matched so that panel a is comparable to d and panel b to panel f). Panels c--f are for main sequence stars of magnitude $r=26$, 25, 24, and 23 respectively in M31, with the appropriate finite source size effect, but without diffraction. The flux perturbation $F_0(A-1)$ is the same for panels c--f, and correspond to a 26th magnitude source in the difference image.}
%\vspace{-70pt}
%\end{figure}

\subsection{Behavior of the microlensing event rate for very low lens masses}
\label{ss:lowmass}

While the microlensing event rates can be computed for any $\mpbh$ using the formalism of \S\ref{ss:muRate}, it is instructive to consider some key scaling relations that can be derived analytically. We will consider these scaling relations in the limit that (i) only nearby lenses ($D_{\rm OL}\ll D_{\rm OS}$) can achieve the required amplification due to finite source size effects; and (ii) the event durations at typical Galactic velocities are $\ll t_{\rm min}$, so that only the few events with the smallest transverse velocities ($v_\perp \ll \bar \sigma$) contribute. Both of these turn out to be the relevant limits for small PBH masses ($\mpbh\lesssim 10^{-10}\,M_\odot$), order-unity amplification cuts, and main sequence source stars in M31.

We focus our discussion of these scaling laws on the geometric case. The reason for this is that the maximum possible amplification $A_{\rm max}(y)$, in the diffractive case, satisfies
\begin{equation}
A_{\rm max}(y)-1 = \frac{2\pi y}{1-e^{-2\pi y}} - 1 = \pi y + \frac{\pi^2}3 y^2 + ...\,,
% q = 2 pi y
% q/(1-e^-q) = q/(q - q^2/2 + q^3/6 - ...)
% = 1/(1 - q/2 + q^2/6 - ...)
% = q/2-q^2/6 + q^2/4 + ... = q/2 + q^2/12 + ...
\end{equation}
where again $y=4\pi G\mpbh/(\lambda c^2)$ depends on the wavelength of observation and the PBH mass. If the threshold $A_{\rm min}$ is above $A_{\rm max}(y)$, then there are no events, and if $A_{\rm min}-1\ll A_{\rm max}(y)-1$, we find that the geometric approximation gives the correct order of magnitude.

Let us first understand our two assumptions. Assumption (i) requires that a lens at $D_{\rm OL}\sim D_{\rm OS}/2$ would have an Einstein radius smaller than the source radius: $R_{\rm E,S} <\xi^{1/2} R_{\rm S}$, where $\xi$ is a dimensionless parameter of order $A_{\rm min}-1$. (We have inserted this parameter because we want our scaling laws to be valid for $A_{\rm min}-1\ll 1$. In the geometric limit, and for large source size, a lens in front of the source will lead to an amplification of $1 + 2R_{\rm E,S}^2/R_{\rm S}^2$.) This leads to
\begin{equation}
\mpbh < m_{\rm PBH,crit} = \frac{\xi R_{\rm S}^2c^2}{4\pi G D_{\rm OS}} \sim 10^{-9} \xi \left( \frac{R_{\rm S}}{R_\odot} \right)^2 \left( \frac{770\,\rm kpc}{D_{\rm OS}} \right)\, M_\odot.
\end{equation}
If $\mpbh \ll m_{\rm PBH,crit}$, then we have $R_{\rm E,S} < \xi^{1/2} R_{\rm S}$ and hence amplification to $A>A_{\rm min}$ is possible if
\begin{equation}
D_{\rm OL} < D_{\rm OL,max} \approx \frac{\mpbh}{m_{\rm PBH,crit}} D_{\rm OS}.
\end{equation}
Assumption (ii) requires that
\begin{equation}
t_{\rm min} > \frac{(D_{\rm OL}/D_{\rm OS})R_{\rm S}}{\sigma}
= \frac{\mpbh}{m_{\rm PBH, crit}} \frac{R_{\rm S}}{\sigma}
= 4100 \frac{\mpbh}{m_{\rm PBH, crit}} \left( \frac{R_{\rm S}}{R_\odot} \right) \left( \frac{170\,\rm km \,s^{-1}}{\sigma} \right)\,{\rm s}.
\end{equation}
For a minimum time of $\sim 400$ s (the appropriate order of magnitude for the HSC survey), our validity criteria will be met for PBHs below 0.1$m_{\rm PBH,crit} \sim 10^{-10}\,M_\odot$.

Once this is achieved, we may estimate the event rate $\Gamma$ by integrating Eq.~(\ref{eq:Gamma-bv}). The maximum impact parameter is $b_{\rm max} = (D_{\rm OL}/D_{\rm OS})R_{\rm S}$ (the requirement for an event is for the lens to transit the source, since the source size dominates over the Einstein radius) and the velocity integral extends up to $v_{\perp,\rm max} = 2(D_{\rm OL}/D_{\rm OS})R_{\rm S}/t_{\rm min}$. We take the limit of small $v_\perp$, so that $P(v_\perp) \approx (v_\perp/\sigma^2)e^{-\bar v_\perp^2/2\sigma^2}$. Then -- doing the $v_\perp$ integral in Eq.~(\ref{eq:Gamma-bv}), and neglecting the $e^{-\bar v_\perp^2/2\sigma^2}$ factor (which makes only a few tens of percents difference and would be a distraction in what follows) -- we get
\begin{eqnarray}
\Gamma &\approx& \int_0^{D_{\rm OL,max}} \frac{2\rho_{\rm DM}(D_{\rm OL})}{3\mpbh}
b_{\rm max}
\frac{v_{\perp,\rm max}^3}{\sigma^2}
 P(v_\perp|d_{\rm OL}) \, {\rm d}d_{\rm OL}
\nonumber \\
&\approx&
\int_0^{D_{\rm OL,max}} \frac{2\rho_{\rm DM}(D_{\rm OL})}{3\mpbh}
\frac{D_{\rm OL}}{D_{\rm OS}}R_{\rm S}
\frac{[2(D_{\rm OL}/D_{\rm OS})R_{\rm S}/t_{\rm min}]^3}{\sigma^2}
\, {\rm d}d_{\rm OL}
\nonumber \\
&\approx&
\int_0^{D_{\rm OL,max}} \frac{8\rho_{\rm DM}(D_{\rm OL})}{3\mpbh}
\frac{D_{\rm OL}^4R_{\rm S}^4}{D_{\rm OS}^4 \sigma^2 t_{\rm min}^3}
\, {\rm d}d_{\rm OL}
\nonumber \\
&\approx&
\frac{8\rho_{\rm DM}(D_{\rm OL,max}) \mpbh^4}{15 m_{\rm PBH,crit}^5}
\frac{R_{\rm S}^4 D_{\rm OS}}{ \sigma^2 t_{\rm min}^3},
\label{eq:Gamma-limit}
\end{eqnarray}
where in the last approximation we evaluated the DM density at $D_{\rm OL,max}$ since the weight in the integrand is sharply peaked there\footnote{It is true that $\rho_{\rm DM}$ decreases as one looks outward through a halo; however as long as the drop-off is shallower than $1/r^5$, one is dominated by distant events.}, and we analytically evaluated $\int_0^{D_{\rm OL,max}} D_{\rm OL}^4 \,{\rm d}D_{\rm OL} = D_{\rm OL,max}^5/5$. 

\begin{figure*}
\begin{subfigure}{.32\textwidth}
\centering
\includegraphics[width=.98\columnwidth]{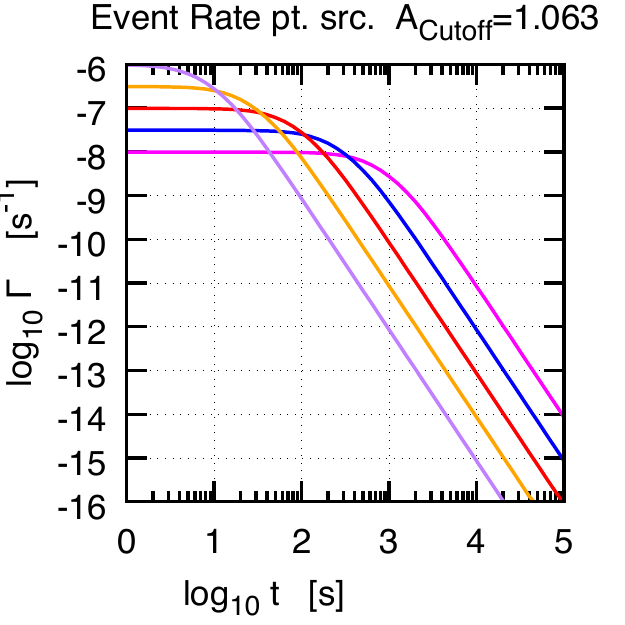}
\end{subfigure}
\begin{subfigure}{.32\textwidth}
\includegraphics[width=.98\columnwidth]{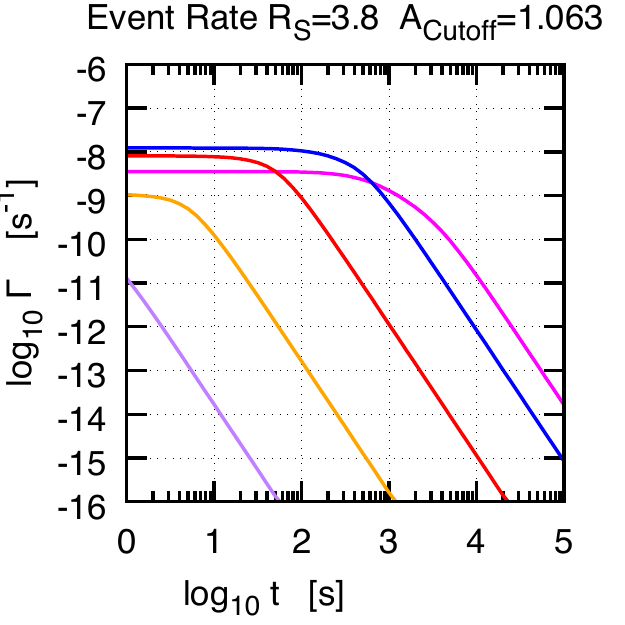}
\end{subfigure}
\begin{subfigure}{.32\textwidth}
\centering
\includegraphics[width=.98\columnwidth]{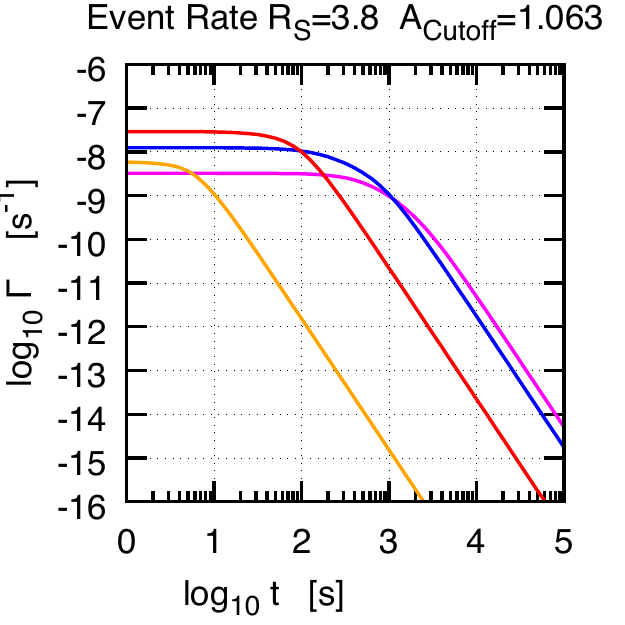}
\end{subfigure}

\begin{subfigure}{.32\textwidth}
\centering
\includegraphics[width=.98\columnwidth]{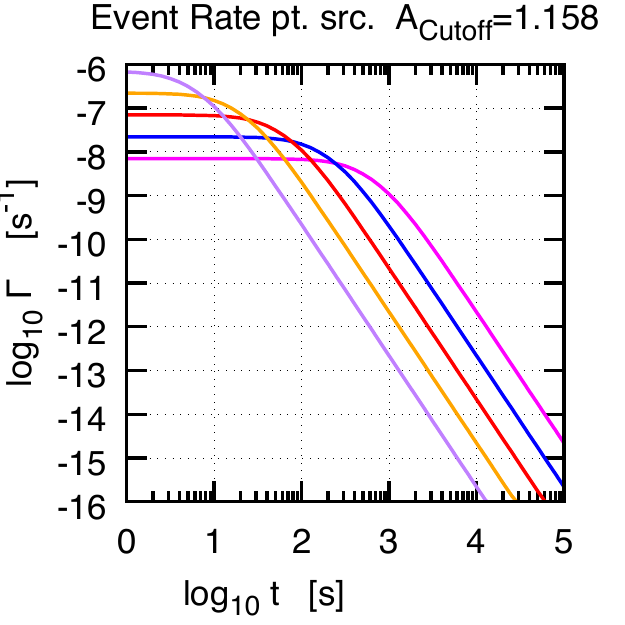}
\end{subfigure}
\centering
\begin{subfigure}{.32\textwidth}
\centering
\includegraphics[width=.98\columnwidth]{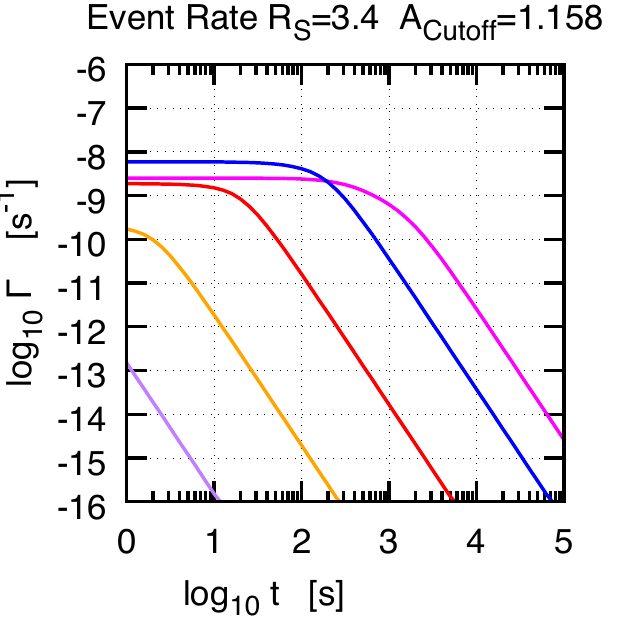}
\end{subfigure}
\centering
\begin{subfigure}{.32\textwidth}
\centering
\includegraphics[width=.98\columnwidth]{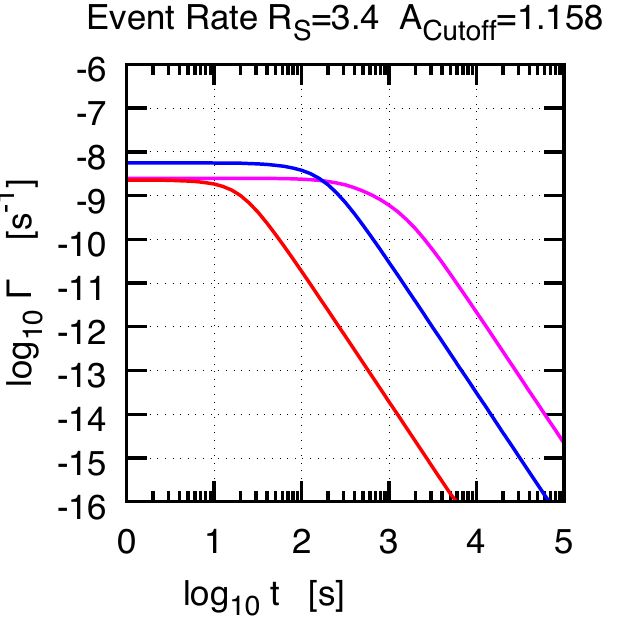}
\end{subfigure}

\begin{subfigure}{.32\textwidth}
\centering
\includegraphics[width=.98\columnwidth]{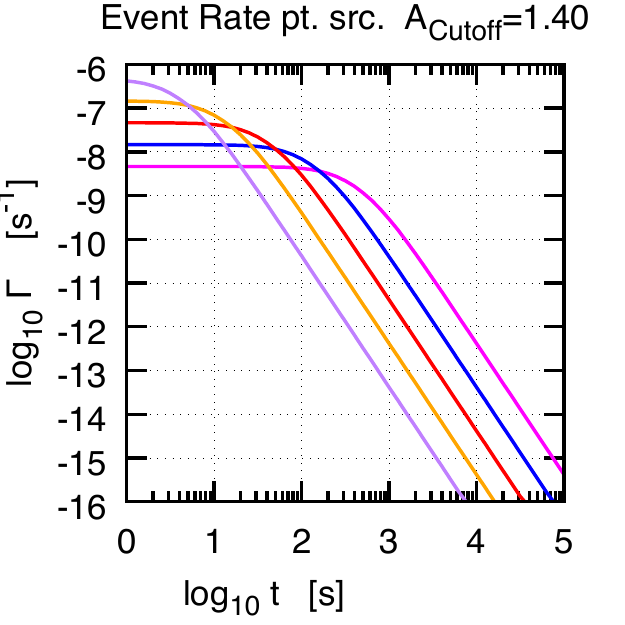}
\end{subfigure}
\centering
\begin{subfigure}{.32\textwidth}
\centering
\includegraphics[width=.98\columnwidth]{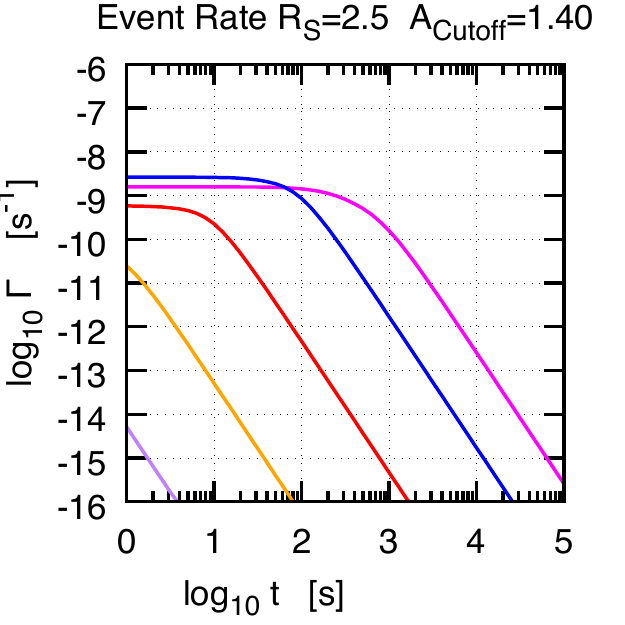}
\end{subfigure}
\centering
\begin{subfigure}{.32\textwidth}
\centering
\includegraphics[width=.98\columnwidth]{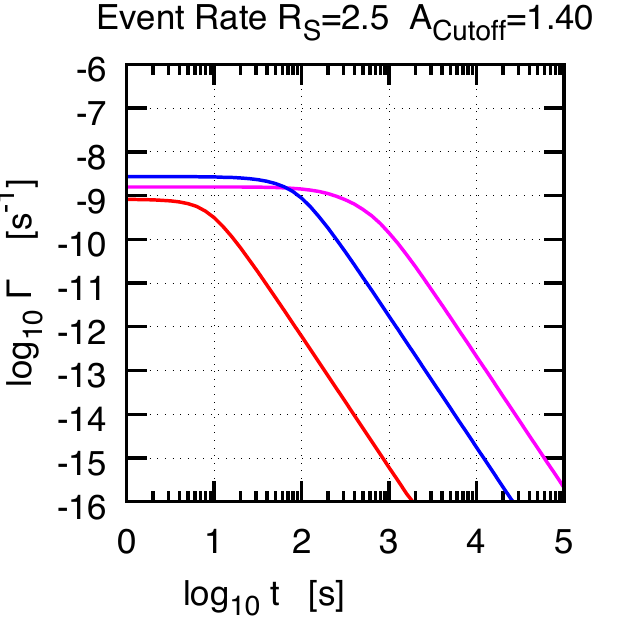}
\end{subfigure}

\begin{subfigure}{.32\textwidth}
\centering
\includegraphics[width=.98\columnwidth]{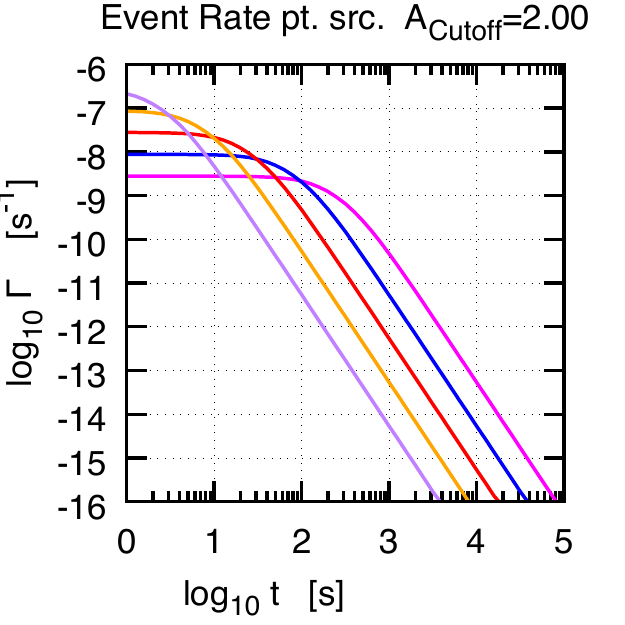}
\end{subfigure}
\centering
\begin{subfigure}{.32\textwidth}
\centering
\includegraphics[width=.98\columnwidth]{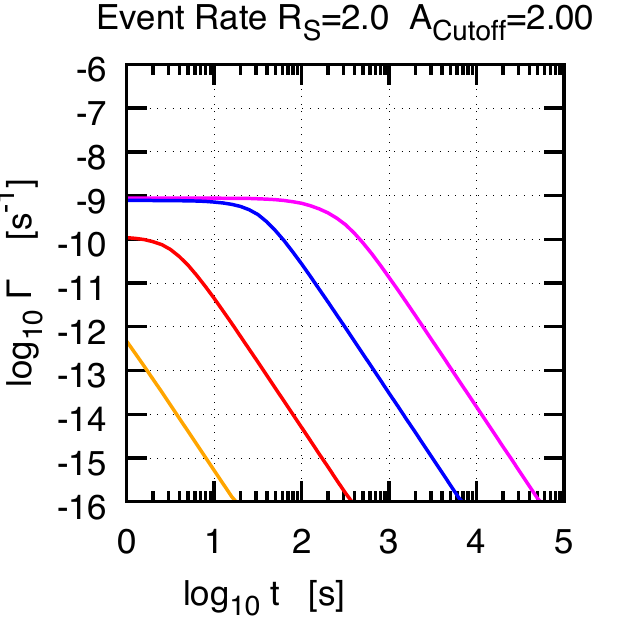}
\end{subfigure}
\centering
\begin{subfigure}{.32\textwidth}
\centering
\includegraphics[width=.98\columnwidth]{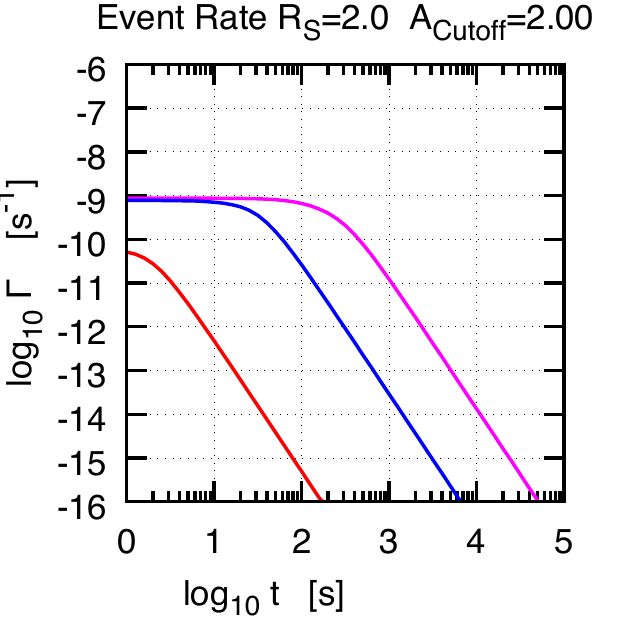}
\end{subfigure}
  \caption{Comparisons of different microlensing event rates. The first column represents point sources with geometric optics, the second column represents finite source size in geometric optics, and the third column represents finite source size with diffractive optics. All figures have the same color code: $\mpbh = 10^{-9} M_\odot$ is magenta, $10^{-10}$ is blue, $10^{-11}$ is red, $10^{-12}$ is orange, and $10^{-13}$ is purple.  In the diffractive case, the rates for sufficiently low $\mpbh$ are zero and hence not shown.}
  \label{fig:deltacomp}
\vspace{-30pt}
\end{figure*}

The key scaling relations seen in \S\ref{ss:muRate} can be understood from Eq.~(\ref{eq:Gamma-limit}). In particular, if we get into the regime where the lenses are so nearby that $\rho_{\rm DM}$ is effectively constant (i.e.\ $\mpbh<0.01m_{\rm PBH,crit}\sim 10^{-11}\xi\,M_\odot$, where $D_{\rm OL}$ is small compared to the distance of the Solar System from the center of the Milky Way), then:
\begin{list}{$\bullet$}{}
\item The rate of events is steeply falling with more stringent duration cuts, $\Gamma\propto t_{\rm min}^{-3}$. (This is because an event that is twice as long requires half the transverse velocity. In 2 dimensions, probability scales as $v_\perp^2$, and the conversion from lens abundance to frequency contains yet another factor of $v_\perp$.)
\item The rate of events is steeply rising with PBH mass, $\Gamma\propto \mpbh^4$. (This is because if the PBH mass is doubled, one has access to lenses twice as far away so there is twice as much path length $\int {\rm d}D_{\rm OL}$. One can then accept lenses with twice the transverse velocity, so we gain a factor of 8 in accordance with the discussion above on $t_{\rm min}$. A further factor of 2 comes from the larger allowable impact parameter to pass in front of the star. But one loses a factor of 2 in lens number density.)
\item The rate of events is steeply falling with source radius, $\Gamma\propto R_{\rm S}^{-6}$, when we take into account that $m_{\rm PBH,crit} \propto R_{\rm S}^2$. (This is because if the source radius is doubled, the lens must be brought a factor of 4 closer to keep the ratio of source size to Einstein radius fixed (so we lose a factor of 4 in path length). The Einstein radius in physical units in the lens plane is then a factor of 2 smaller, so we lose a factor of $2^3=8$ from the transverse velocity factors and a factor of 2 in maximum impact parameter.)
\item The rate of events scales with $\xi^{-5}\propto (A_{\rm min}-1)^{-5}$ for small $A_{\rm min}$ (due to the $m_{\rm PBH,crit}$ scaling). This is because if $A_{\rm min}-1$ is reduced by a factor of 2, we can achieve the desired amplification by placing the lens twice as far away; this means at any given time there are 8 times as many candidate lenses in front of the star. We can also accept lenses that move twice as fast, so we increase the area in 2D ${\mathbf v}_\perp$-space by a factor of 4. This leads to $8\times 4=32$ times as many events.
\end{list}
As a simple check of these scaling relations, one predicts that at low $\mpbh$, the $R_S=3.8R_\odot$, $A_{\rm min}=1.063$ curve should be a factor of $(3.4/3.8)^6 (0.158/0.063)^5 = 50$ times larger than the $R_S=3.4R_\odot$, $A_{\rm min}=1.158$ curve in Fig.~\ref{fig:deltacomp}. Comparing the top two panels in the middle column of the figure shows good agreement with this expectation.

\subsection{Discussion and future prospects}

The HSC M31 survey is estimated to have included $8.7\times 10^7$ stars down to magnitude $r=26$, observed for 7 hours, and with a cut requiring 3 consecutive exposures (2 minutes each: $t_{\rm min} \sim 360$ s) with the flux in the difference image above $5\sigma$ (the threshold varies throughout the observing period, but goes down to $r\sim 26$). For this duration, there is a $>5\%$ probability of no events if $\Gamma\le (-\ln0.05)/(8.7\times 10^7)/(7\,{\rm hr}) = 1.4\times 10^{-12}\,{\rm s}^{-1}$. Based on the rates in Figure~\ref{fig:deltacomp}, we can see that for $r=24$--26 sources (bottom three rows) the rate is far below this for $\mpbh\lesssim 10^{-11}\,M_\odot$, and thus a search of 7 hour duration will not be sensitive to such low masses.\footnote{There is a rate exceeding $1.4\times 10^{-12}$ s$^{-1}$ for $r=23$ sources; however the number of sources at $r=23$ is more than an order of magnitude lower, and many of them may be giants.} These expectations are consistent with the reported limits from HSC \cite{2019NatAs.tmp..238N}.

The source radius scaling in particular should indicate how important it is to have an accurate estimate of the source size.\footnote{The scaling with $t_{\rm min}$ is visible in Fig.~24 of Ref.~\cite{2019NatAs.tmp..238N}; the slope there is $-4$ instead of $-3$ since the event rates are differential rather than integral.} Indeed, since Ref.~\cite{2019NatAs.tmp..238N} used $R_{\rm S} = R_\odot$, whereas a 26th magnitude main sequence star in M31 has radius $2.0R_\odot$, it is possible that in the $\sim 10^{-11}M_\odot$ regime the limits on $\fpbh$ in Ref.~\cite{2019NatAs.tmp..238N} are still too optimistic by a factor of $\sim 2^6\sim 64$. However, this has little impact on the range of $\mpbh$ that is constrained, since the constraint curve in the $(\mpbh,\fpbh)$ plane is nearly vertical.

%As reported by Ref.~\cite{2019NatAs.tmp..238N}, the PBH limits from the HSC M31 survey are uncertain below $\sim 10^{-10}\,M_\odot$; while a more detailed analysis would be required to derive a complete new constraint curve, we can conclude that PBH masses in the $\le 10^{-11}\,M_\odot$ range are ``safe'' from this constraint.

Unfortunately, the scaling laws discussed here suggest that it will be hard to push to much lower $\mpbh$ than the current HSC constraint, even with large increases in telescope time. Because of diffraction effects, at $r2$-band achieving amplification $A=(1.063,1.158,1.4,2)$ requires $\mpbh>(0.7,1.6,4,9)\times 10^{-12}M_\odot$. If one wants to push down a factor of 2 in $\mpbh$, we require a factor of 2 in $A_{\rm min}-1$; based on the scaling relations in \S\ref{ss:lowmass}, the event rate at fixed $t_{\rm min}$ is actually enhanced by a factor of 2 (there is a factor of 16 loss from $\Gamma\propto\mpbh^4$ and a factor of 32 gain from $\Gamma\propto\xi^{-5}$). However, there is a challenge in detecting these events; for example, the HSC ETC\footnote{See {\tt https://hscq.naoj.hawaii.edu/cgi-bin/HSC\_ETC/hsc\_etc.cgi}. We used a lunar age of 1 day, 0.7 arcsec seeing, and the default aperture of 2 arcsec diameter. Note, however, that matched filter photometry would achieve slightly higher S/N.} estimates that in 6 minutes of live time, one achieves S/N=60 (so $4\sigma$ detection of an $A=1.063$ event) on a 23th magnitude point source in $r2$-band. If we make the microlensing signal a factor of 2 smaller, we must increase $t_{\rm min}$ by a factor of 4 to achieve the same threshold S/N, which reduces the event rate by a factor of $4^3=64$. Thus, once we enter the diffractive regime, each factor of 2 improvement in reach for $\mpbh$ implies a factor of 32 reduction in the event rate and hence a factor of 32 increase in observing time to test the $\fpbh=1$ hypothesis (even if there are no false positives).
%The existing HSC constraint comes from one night of observations; a factor of 32 increase in observing time to push the constraints down by a factor of 2 to $2\times 10^{-12}M_\odot$ may be possible, but a factor of 1024 (i.e., 1000 nights of M31 observations) to reach $10^{-12}M_\odot$ is not realistic.

Another way to reach toward lower $\mpbh$ at fixed telescope time is to build a better ``light bucket'' and collect more photons from the source stars. Since $t_{\rm min}$ scales as the inverse of the source count rate at fixed $A_{\rm min}$, the time $t_{\rm obs}$ required to constrain $\fpbh=1$ at fixed $\mpbh$ will scale as the inverse cube of $Q$, the $(S/N)^2$ achieved per unit time: $t_{\rm obs}\propto Q^{-3}$ Note that $Q$ is proportional to collecting area so long as the seeing and hence source-to-background ratio remain fixed. A larger aperture telescope would help, but there is no wide-field camera analogous to HSC planned for the future large telescopes. Since the source stars that dominate the constraint are blue (B or early A), some marginal improvement may be obtained by switching to the $g$-band, which increases $Q$ by a factor of $\sim 2$ (i.e., equivalent to doubling the light bucket power) according to the HSC ETC, or possibly introducing a wide ($g+r$) filter. However, because $t_{\rm obs}$ scales with the inverse cube of $\dot N$ but with the 5th power of $\mpbh$, even doubling the source count rate gains one a factor of only $2^{3/5}=1.5$ in $\mpbh$-space.

One could also improve the $(S/N)^2$ per unit time by taking advantage of the higher resolution from space. As a simple example, we might consider a survey of (part of) M31 with WFIRST.\footnote{We use the WFIRST ETC \cite{2012arXiv1204.5151H}, version 16, with the Cycle 7 reference information on the Project website, {\tt https://wfirst.gsfc.nasa.gov/science/WFIRST\_Reference\_Information.html}.} We consider the $R$ band, which is the bluest band in WFIRST and has a similar central wavelength to HSC $r2$. Assuming $3\times 2$ minute exposures (with small dithers to avoid false positives from unstable pixels), and applying an added background of 22 mag arcsec$^{-2}$ from M31 itself (a typical value; see Fig.~2 of Ref.~\cite{2006MNRAS.365.1099K}), we find a $S/N$ of 84 vs.\ 56 (for WFIRST vs.\ HSC) for an $r=23$ source, and 8.2 vs.\ 3.6 for an $r=26$ source. The advantage in effective light bucket power $Q$ is 2.2 for the $r=23$ case and 5.2 for the $r=26$ case. (By resolving out more stars, WFIRST would actually have a lower surface brightness from blended stars, so these estimates may somewhat underestimate its advantage; however even if the higher resolution eliminated blending entirely these $(S/N)^2$ advantages would only improve to 2.7 and 9.0.) There are also differences in the number of source-hours that can be covered per day of wall clock time: WFIRST observes 24 hours per day, and while it has a $7\times$ smaller field of view, about half of the HSC field lies outside the star-forming parts of M31 that would host massive main sequence source stars; we would expect these three effects to roughly cancel. Since the number of source-hours of total exposure required to achieve a given $\mpbh$ constraint scales as $Q^{-3}$, the space-based survey option should be investigated further and a careful forecast performed to determine its reach in parameter space. However, we do not expect to come anywhere close to closing the low-mass window in $\mpbh$ by this method.

Finally, one could change $D_{\rm OS}$, and consider a source galaxy more distant than M31. Taking into account the scaling of $m_{\rm PBH,crit}$, we see that $\Gamma \propto D_{\rm OS}^6$. However, by placing a source farther away, the count rate is reduced as $\propto D_{\rm OS}^{-2}$. This results in $t_{\rm min}\propto D_{\rm OS}^2$ or $D_{\rm OS}^4$ (for source- or background-dominated cases), and with the $t_{\rm min}^{-3}$ scaling we would have the event rate $\Gamma\propto D_{\rm OS}^0$ or $D_{\rm OS}^{-6}$ (again for source- or background-dominated cases). The $23<r<26$ stars that dominate the M31 constraint are near the boundary; putting them farther away thus will not help.

These scaling laws for pushing to lower masses stand in contrast to the situation for improving constraints on $\fpbh$ in the planetary-to-stellar mass range. Here the main challenge is the ``background'' of real microlensing events from astrophysical compact objects \cite{2019PhRvD..99h3503N}; for example, the WFIRST Galactic Bulge survey is expected to yield $\sim 5\times 10^4$ astrophysical microlensing events \cite{2019ApJS..241....3P}. However with large statistical samples, distributions of event parameters may enable PBH models to be distinguished (e.g., \cite{2019RNAAS...3d..58L}).

We conclude that the M31 microlensing technique -- while currently the leading constraint on PBHs in a wide range of masses -- is unlikely to probe to $\ll 10^{-12}M_\odot$ in the foreseeable future. This is consistent with the conclusions recently discussed in Ref.~\cite{2019arXiv190506066S}. Thus other types of observations will be necessary to fully explore the asteroid-mass PBH window. One interesting suggestion is to do microlensing studies in the X-ray \cite{2018arXiv181201427B}, although the required observations may be some time in the future. The alternative is to consider the astrophysical effects of low-mass PBHs.

\section{Dynamical capture of PBHs in stars}
\label{sec:capture}

Another class of potential constraints comes from stars capturing a PBH by dynamical friction. In most of the parameter space, the aftermath is the star gets destroyed: the PBH settles into the center of the star, then the PBH grows by Bondi accretion. Finally then the star may be eaten by accretion onto the black hole, or possibly feedback from such accretion might disrupt it. The probability of capturing the PBH is higher in denser regions with lower velocity dispersions, which is why some papers focus on globular clusters \citep{2013PhRvD..87b3507C,2013PhRvD..87l3524C}, though these may not have significant dark matter content. We also consider the more general case of stars in the Milky Way Galaxy. We argue that {\em survival} of stars (in the sense of requiring the probability per Hubble time of destroying any given star to be $\ll 1$) does not constrain PBHs. However, some stellar destruction events should occur; we estimate their rate and argue that modeling them is a promising avenue toward constraining asteroid-mass PBHs in the future.

\subsection{General considerations on PBH capture}

The problem of PBH capture by a star is commonly treated as a multi-step process (e.g., \cite{2013PhRvD..87l3524C}). First, the PBH passes through the star, loses energy by dynamical friction\footnote{There have been suggestions that the actual energy loss is much greater than given by the dynamical friction formula when the sum over modes is taken into account \cite{2014JCAP...06..026P}, due to generation of surface waves. See, however, Ref.~\cite{2014PhRvD..90j3522D}, who find a much smaller surface wave contribution. We have done our own derivation of Eq.~(13) of Ref.~\cite{2014PhRvD..90j3522D} using Fourier (definite $k_x$ and $k_y$) instead of cylindrical modes, and find the same result.}, and then is captured onto a long elliptical orbit. Then on subsequent periastron passages, the PBH loses additional energy and lowers its apoastron. After enough passages, the PBH orbit becomes completely engulfed in the star, and it settles to the star's center. In this picture, each passage through a star of mass $M$, radius $R$, characteristic density $\rho_\star= M/(\frac43\pi R^3)$, characteristic velocity $v_0=(GM/R)^{1/2}$, and dynamical time $t_{\rm dyn}=R/v_0$ results in a loss of energy of order
\begin{equation}
|\Delta E| \sim F_{\rm drag}R \sim \frac{(G\mpbh)^2\rho_\star}{v_0^2}\ln\Lambda \times R \sim \frac{\mpbh^2}Mv_0^2\ln\Lambda,
\label{eq:DE}
\end{equation}
where $\ln\Lambda$ is the Coulomb logarithm. For a PBH captured from a near-parabolic orbit, the semi-major axis of the $N$th bound orbit is obtained from
\begin{equation}
\frac{GM\mpbh}a \sim N|\Delta E|\sim N\frac{\mpbh^2}Mv_0^2\ln\Lambda ~~~~\rightarrow~~~~
a\sim \frac {MR}{N\mpbh\ln\Lambda}.
\label{eq:aa}
\end{equation}
Since the orbital period scales as $\sim a^{3/2}\sim N^{-3/2}$, and $\sum_{N=1}^\infty N^{-3/2}$ converges, the first few orbits dominate the time required for a PBH to sink into the star. Therefore the capture is typically viewed as governed by two processes: (i) the initial encounter, which sets the rate of captures (determined by the abundance of slow-moving PBHs); and (ii) whether the initial ($N=1$) capture orbit period is short enough that the capture proceeds in a timescale of interest (e.g.\ the Hubble time, or the lifetime of the star in question).

This physical picture of PBH capture is, however, not applicable in a large portion of the parameter space of interest. There are two competing effects that may make the rate of stellar destruction either lower or higher than predicted based on the capture arguments. First, for light PBHs the initial capture orbit will not remain undisturbed since a realistic star is not in an isolated system with Keplerian orbits. If an external perturber (a companion star, or the bulk tidal field of a star cluster or galaxy) is sufficiently strong, the PBH is then pulled into a long eccentric orbit around the target. It will then undergo complex multi-body dynamics; if the PBH is subsequently ejected from the system, or undergoes further orbital evolution such that in the lifetime of the Universe it does not re-contact the star, then it will not be captured into the star's core and does not contribute to constraints from stellar survival. On the other hand, during the process of star formation itself, the star had a time-dependent gravitational potential, and thus even without dynamical friction the energy of passing PBHs is not conserved; thus they may be captured even in the test particle limit. This enhances the rate of stellar destruction \cite{2013PhRvD..87b3507C, 2014PhRvD..90h3507C}.

Let us first show by order-of-magnitude arguments that external perturbers are relevant in a large portion of the parameter space. The orbital timescale (period divided by $2\pi$) for the capture orbit is $\bar P = t_{\rm dyn} (a/R)^{3/2}$. In an external tidal field with gravity gradient $\sim t_{\rm tidal}^{-2}$, this means that on its first orbit, the PBH will acquire specific angular momentum
\begin{equation}
L \sim t_{\rm tidal}^{-2} a^2 \bar P \sim \frac{t_{\rm dyn}R^2}{t_{\rm tidal}^2} \left( \frac aR\right)^{7/2}
\end{equation}
But there is a critical angular momentum $L_{\rm crit}\sim Rv_0 \sim R^2/t_{\rm dyn}$ above which the periastron of the PBH's orbit is outside the star; thus we see that following capture onto a long eccentric orbit, the PBH misses the star on the next periastron if
\begin{equation}
\frac{t_{\rm dyn}R^2}{t_{\rm tidal}^2} \left( \frac aR\right)^{7/2} \gtrsim \frac{R^2}{t_{\rm dyn}}
~~\leftrightarrow~~
\frac aR \gtrsim \left( \frac{t_{\rm tidal}}{t_{\rm dyn}} \right)^{4/7}
~~\leftrightarrow~~
\mpbh \lesssim \frac{M}{\ln\Lambda} \left( \frac{t_{\rm dyn}}{t_{\rm tidal}} \right)^{4/7}.
\end{equation}
As one can see, even for the extreme case of a neutron star ($t_{\rm dyn} \sim 10^{-4}$ s) in the Milky Way disc ($t_{\rm tidal} \sim 10^{15}$ s), this order-of-magnitude estimate suggests that PBHs with mass below a few$\times 10^{-13}\,M_\odot$ will have their capture dynamics substantially modified by external perturbers. For less compact stars (white dwarf or main sequence), dense stellar environments (globular clusters), and multiple star systems, PBHs with even higher masses will be affected by external perturbers.

In order to proceed, one might at first think that an expensive suite of numerical simulations is necessary to determine for what range of PBH masses and for what targets a PBH is likely to be captured in the target star. Fortunately, a simple and useful bound can be derived analytically using phase space arguments, which are valid regardless of the details of the perturbers or their evolutionary history (which may be quite complex if we consider constraints from millisecond pulsar survival).

\subsection{Phase space arguments}

Our starting point is the phase space density $f({\boldsymbol r}, {\boldsymbol v}, t)$ of PBHs -- defined here as the mass of PBHs per unit volume in position space per unit volume in velocity space (units: kg m$^{-6}$ s$^3$).\footnote{This formulation of phase space density is appropriate for describing dark matter, as the phase space density in an astrophysical environment can be computed without needing the particle mass.} We suppose that the PBHs are light compared to the target stars, and that they move in a background of some density $\rho({\boldsymbol r})$, which is the target star.

If particles move according to some evolution equations for ${\boldsymbol r}$ and ${\boldsymbol v}$, then the convective derivative of the phase space density is
\begin{equation}
\frac{{\rm d}f}{{\rm d}t} = -(\nabla_{\boldsymbol r}\cdot\dot{\boldsymbol r} +\nabla_{\boldsymbol v}\cdot\dot{\boldsymbol v})f,
\label{eq:dfdt}
\end{equation}
where the divergences are taken with respect to the stated vectorial argument. For conservative forces, the right-hand side is zero by Liouville's theorem. In the presence of dynamical friction, however, we have the Chandrasekhar drag term:
\begin{equation}
\dot{\boldsymbol v}_{\rm Ch} = -4\pi G^2\mpbh \rho \frac{\ln \Lambda}{v^2}\hat{\boldsymbol v},
\label{eq:dynCh}
\end{equation}
where $\hat{\boldsymbol v}$ is the unit vector in the direction of ${\boldsymbol v}$, $\rho$ is the density of background material, and $\ln\Lambda$ is the Coulomb logarithm. The Chandrasekhar formula as written in Eq.~(\ref{eq:dynCh}) originally applied to a background of slowly moving collisionless particles, however it is also valid for a fluid if the PBH moves supersonically (as it would if passing through a star). Plugging this into Eq.~(\ref{eq:dfdt}), and using the divergence formula $\nabla_{\boldsymbol v}\cdot\dot{\boldsymbol v} = v^{-2} \partial_v (v^2 \dot v)$ valid for $\dot{\boldsymbol v} \parallel {\boldsymbol v}$ (drag only, no lift), we find
\begin{equation}
\frac{{\rm d}f}{{\rm d}t} = \frac{4\pi G^2\mpbh \rho}{v^3} \frac{\partial \ln\Lambda}{\partial\ln v} \,f.
\label{eq:dfdtCh}
\end{equation}
Because $f$ and ${\rm d}f/{\rm d}t$ are invariant under changes of reference frame, Eq.~(\ref{eq:dfdtCh}) remains valid if the star is moving (e.g., it is in a binary or multiple system) so long as we interpret $\ln\Lambda$ and $\ln v$ to be computed in the rest frame of that star. We thus see that if $\ln\Lambda$ is constant -- which is often taken as a first approximation -- then $f$ is conserved along a trajectory even though the force is dissipative. (We will address varying $\ln\Lambda$ below and conclude that it does not substantially change the result.)

One might wonder how the star can capture black holes if $f$ is conserved. The answer is that Eq.~(\ref{eq:dynCh}) and hence Eq.~(\ref{eq:dfdtCh}) do not apply to subsonic PBHs that are already captured. We instead define an ``interior region'' ${\cal C}\subset{\mathbb R}^6$ in phase space, for particles that are on orbits fully inside the star, and an ``exterior region'' ${\mathbb R}^6\setminus{\cal C}$ for orbits that partially pass through the star and are unbound. Then Eq.~(\ref{eq:dfdtCh}) should apply in ${\mathbb R}^6\setminus{\cal C}$. We can then define a PBH capture rate as the flux of PBHs across the boundary $\partial{\cal C}$. The number of PBHs per unit position-space volume per unit time crossing into ${\cal C}$ due to dynamical friction is
\begin{equation}
\frac{{\rm d}\dot N}{{\rm d}^3{\boldsymbol r}} = - \frac1{\mpbh} \oint_{\partial\cal C}  \dot{\boldsymbol v}_{\rm Ch}\cdot {\rm d}{\boldsymbol a}_{\boldsymbol v},
\end{equation}
where ${\rm d}{\boldsymbol a}_{\boldsymbol v}$ is the area element of the 2D boundary of ${\cal C}$ (as described in 3D velocity space). If at position ${\boldsymbol r}$ there is a maximum velocity $v_{\rm max}({\boldsymbol r})$ for a particle orbit to be fully inside the star, then this area element can be written using the differential of solid angle as $v_{\rm max}^2\,d^2\hat{\boldsymbol v}$. Using Eq.~(\ref{eq:dynCh}), this results in a simplification to
\begin{equation}
\frac{{\rm d}\dot N}{{\rm d}^3{\boldsymbol r}} = 4\pi G^2\rho \oint_{S^2} f({\boldsymbol r}, {\boldsymbol v}) \ln\Lambda
{\rm d}^2\hat{\boldsymbol v},
\end{equation}
where the velocity is evaluated at $|{\boldsymbol v}|=v_{\rm max}({\boldsymbol r})$. Integrating over position gives
\begin{equation}
\dot N = 4\pi G^2 \int {\rm d}^3{\boldsymbol r} \,
\rho({\boldsymbol r})
\oint_{S^2} f({\boldsymbol r}, {\boldsymbol v}) \ln\Lambda\,
{\rm d}^2\hat{\boldsymbol v},
\label{eq:Ntemp}
\end{equation}
where the outer (3D) integral is taken over all positions, and the inner (2D) integral is taken over the directions in velocity space $v=v_{\rm max}$. The integrand always contributes positively to $\dot N$; there is a net positive rate at which PBHs are eaten by the star. Their phase space density in the exterior region is conserved if $\ln\Lambda$ is constant, and a pileup in phase space must occur somewhere in the interior region.

Using the fact that the solid angle of the sphere is $4\pi$ to do the inner integral in Eq.~(\ref{eq:Ntemp}), we find
\begin{equation}
\dot N \le   \int {\rm d}^3{\boldsymbol r}\,
16\pi^2 G^2 \rho f_{\rm max} \ln \Lambda = 16\pi^2 G^2 M f_{\rm max} \ln \Lambda,
\label{eq:Ndot}
\end{equation}
where $M = \int \rho({\boldsymbol r})\,{\rm d}^3{\boldsymbol r}$ is the mass of the target and $f_{\rm max}$ is the maximum phase space density of PBHs. We thus see that Eq.~(\ref{eq:Ndot}) provides an upper bound on the rate of capture of PBHs. If the trajectories that arrive at $\partial{\cal C}$ all came from near zero velocity if traced backward to their initial encounter, and $f$ is maximum at zero velocity, then $f$ in Eq.~(\ref{eq:Ntemp}) is constant and the bound is saturated.

In the case of PBH dark matter of density $\rho_{\rm PBH}$ and a Gaussian velocity dispersion $\sigma$, we have $f_{\rm max}=\rho_{\rm PBH}/(2\pi\sigma^2)^{3/2}$ and then the bound on the mean number $N$ of captured PBHs in a star of mass $M$ after time $t$ becomes
\begin{eqnarray}
N \le N_{\rm max} &=& 4\sqrt{2\pi}\, \frac{G^2 M\rho_{\rm PBH}t}{\sigma^3} \ln \Lambda 
\nonumber \\
&=& 1.87\times 10^{-7} \,\frac{M}{M_\odot}\, \frac{\rho_{\rm PBH}c^2}{1\,{\rm GeV}\,{\rm cm}^{-3}}\,\frac{t}{10\,\rm Gyr}\, \left(\frac{\sigma}{200\,{\rm km}\,{\rm s}^{-1}}\right)^{-3} \,\frac{\ln\Lambda}{30}.
\label{eq:NdotB}
\end{eqnarray}

We consider two subtle points in the PBH capture rate in Appendix~\ref{app:subtleties}: (i) the fact that $\ln\Lambda$ is not a constant; and (ii) the fact that due to tidal excitation of stellar oscillations, a PBH passing near a star can lose energy even though the local density of matter it encounters has $\rho=0$ (this is really a deficiency of the Chandrasekhar formula, which assumes a constant background density). We argue there that these both lead to only small corrections to the capture rate, Eq.~(\ref{eq:Ndot}).

It is interesting to note that the bound on the capture rate of PBHs in Eq.~(\ref{eq:Ndot}) does not depend on $\mpbh$: more massive PBHs experience more dynamical friction, but there are fewer of them (assuming PBHs make up a fixed fraction of the dark matter). It also depends only on the mass of the target star; the details of the internal structure are not relevant. The phase space conservation arguments are equally applicable if the star is in a binary, globular cluster, or there is some other external gravitational perturbation. All of this makes it especially useful for determining whether survival of a particular type of star in a particular environment can potentially provide a useful constraint on PBHs.

Finally, we note that Eq.~(\ref{eq:NdotB}) is an inequality. However, in the special case that the star is moving through the DM halo at velocity ${\boldsymbol v}_0$, and nearly all of the PBH capture trajectories if traced back came from small velocity relative to the star ($|{\boldsymbol v}-{\boldsymbol v}_0|\ll\sigma$), then the phase space density on these trajectories is $f_{\rm max} e^{-v_0^2/2\sigma^2}$ and hence $N = N_{\rm max} e^{-v_0^2/2\sigma^2}$. This is likely to be the case for isolated main sequence stars, which (for $\mpbh\ll M$) can only capture PBHs from initial velocities that are small compared to the escape velocity from the star, and for which the energy loss $|\Delta E|$ (from Eq.~\ref{eq:DE}) is easily seen to be $\ll \mpbh \sigma^2/2$ for all masses considered in this paper ($\mpbh<10^{-7}\,M_\odot$). It would also be expected to be the case for most binary stars in our Galaxy, whose orbital velocities (and hence the typical velocity of PBHs that would be dynamically captured) are $\ll \sigma$. Therefore -- while we must remember that Eq.~(\ref{eq:NdotB}) is only an upper bound on the number of PBH captures -- in practical situations like stars in the Milky Way, it is probably an overestimate by at most a factor of a few.

\subsection{Post-capture dynamics}
\label{ss:postcapture}

Thus far, we have computed a ``capture'' rate, which is the rate at which PBHs cross into a 6D region ${\cal C}$ in phase space (defined as orbits that are fully inside the star). We have not yet determined how long it takes the PBH to sink to the center of the star. To do this, we recall that the Chandrasekhar formula has a drag time
\begin{equation}
t_{\rm drag} = \frac{v^3}{4\pi G^2\mpbh \rho \ln\Lambda}.
\end{equation}
%Once a star has been captured, we take as an order of magnitude $v\sim v_0$ and $\rho\sim\rho_\star$, which then gives:
%\begin{equation}
%t_{\rm drag} \sim \frac{v_0^3}{4\pi G^2\mpbh \rho_\star \ln\Lambda}
%= \frac{v_0^3R^3}{3 G^2\mpbh M \ln\Lambda}
%= \frac{Mt_{\rm dyn} }{3 \mpbh \ln\Lambda}.
%\end{equation}
A PBH inside a star will also accrete matter. In the absence of feedback, and once the PBH slows to subsonic speeds, we expect it to accrete at the Bondi rate $\dot m_{\rm PBH} = \mpbh/t_{\rm Bondi}$ \cite{1952MNRAS.112..195B}, where
\begin{equation}
t_{\rm Bondi} = \frac{c_s^3}{4\pi \lambda G^2 \mpbh \rho}
\label{eq:tBondi}
\end{equation}
and $\lambda$ is a dimensionless parameter that depends on the equation of state of the stellar matter ($\lambda = 0.25$ for adiabatic index $\gamma=\frac53$, as expected in most main sequence stellar interiors). Since we would typically expect $v\sim c_s$ and $\ln\Lambda\approx 30$, we will have $t_{\rm drag} < t_{\rm Bondi}$. Since $t_{\rm Bondi}\propto 1/\mpbh$, the differential equation for the evolution of the PBH mass can be integrated easily; it will follow
\begin{equation}
\mpbh = \frac{m_{\rm PBH,init}}{1-t/t_{\rm Bondi,init}}
\end{equation}
so long as we neglect the evolution of the star. There is a runaway at $t \rightarrow t_{\rm Bondi,init}$. We will return to what happens then in \S\ref{ss:runaway}.

If $t_{\rm Bondi}$ is short compared to the age of the Universe (or becomes short in the stellar remnant, which usually has larger $\rho/c_s^3$), then we expect that the PBH will destroy its host star or stellar remnant. Conversely, if $t_{\rm Bondi}$ is too long, then the PBH remains very small and an external observer will not notice anything unusual. We note that if we scale to parameters relevant to the Sun's core \cite{2001ApJ...555..990B}:
\begin{equation}
t_{\rm Bondi} = 9.6 \left( \frac{0.25}{\lambda} \right) \left( \frac{c_s}{505\,\rm km\,s^{-1}} \right)^3
\left( \frac{152\,\rm g\,cm^{-3}}{\rho} \right)
\left( \frac{10^{-16}\,M_\odot}{\mpbh} \right)\,{\rm Gyr}.
\end{equation}
Thus in Sun-like stars, capture of a PBH of mass exceeding a few$\times 10^{-16}\,M_\odot$ will result in the star being destroyed in less than the age of the Galaxy or the main sequence lifetime. The minimum PBH mass required for Bondi accretion in the age of the galaxy scales as $\propto T^{3/2}/\rho$, so it is actually lower for lower-mass stars. By interpolating the zero-age main sequence models at composition $Y=0.25$ and $Z=0.01$ \cite{1983ApJ...266..747V}, we find that for main sequence stars at $M<0.20M_\odot$, the Bondi time is $<10$ Gyr for $\mpbh>3.5\times 10^{-17}M_\odot$, and thus for all PBH masses allowed by evaporation constraints \cite{2010PhRvD..81j4019C}. The Bondi time is also short for white dwarfs, if we use the $c_s$ and $\rho$ that correspond to the central density of the white dwarf sequence \cite{1986bhwd.book.....S}.

%For a Sun-like star ($t_{\rm dyn} = 1400$ s) and $\ln\Lambda = 30$, we see that $t_{\rm drag} \sim 10\,$Gyr (the lifetime of the Sun) for $\mpbh \sim 5\times 10^{-17}\,M_\odot$. For PBH masses much larger than this, the PBH should sink to the center of the Sun. The minimum mass will be larger .

The lifetime of the star could conceivably be extended if the accretion rate were reduced to $\dot\mpbh\ll \dot m_{\rm Bondi}$. It has been proposed \cite{2009arXiv0901.1093R} that this could occur due to radiation pressure: the accretion luminosity $L = \epsilon\dot\mpbh c^2$ will limit accretion to the Eddington rate
\begin{equation}
\dot m_{\rm Edd} = \frac{4\pi G\mpbh}{\epsilon \kappa c}.
\end{equation}
This is valid {\em if} the accretion luminosity is transported by radiation. However, the Eddington limit is really a limit only to the luminosity carried by radiation, $L_{\rm rad}$. In a very optically thick stellar core, it is possible for energy to be carried primarily by convection instead. This happens in the cores of massive main-sequence stars, where energy production is centrally concentrated and the total luminosity at a given radius $L(r)$ can exceed the Eddington luminosity $L_{\rm Edd}$ computed based on the enclosed mass $m(r)$ (see Ref.~\cite{2015A&A...580A..20S} for a detailed discussion). Thus in the presence of convection, we may have $L_{\rm rad}\ll \epsilon \dot m_{\rm PBH} c^2$, and the ``Eddington limit'' does not represent a limit to the PBH’s mass growth rate. Even if such a limit did apply, however, the growth timescale $\mpbh/\dot m_{\rm Edd}$ is short compared to the age of the Galaxy and of the stars that make up the bulk of the stellar mass. Thus radiative feedback would not be expected to ``save'' a star from destruction by a PBH trapped in its core.

Another way that $\dot m_{\rm PBH}$ could be reduced would be if the Hawking radiation from the black hole were energetically significant on the scale of the accretion flow. The relevant dimensionless ratio is the ratio of the ``visible'' Hawking luminosity $L_{\rm H,vis}$ (i.e., excluding neutrinos and gravitons that will not interact with the accreting material) to the thermal energy of the accreting mass, $ \dot m_{\rm Bondi} u_{\rm init}$, where $u_{\rm init} = c_s^2/[\gamma(\gamma-1)]$ is the energy density per unit mass. The Hawking luminosity is $C_{\rm H}f_{\rm vis}c^2/\mpbh^2$, where $C_{\rm H} = 5.34\times 10^{25}$ g$^3$ s$^{-1}$ and $f_{\rm vis}$ is a dimensionless factor describing the effective number of visible degrees of freedom that can be emitted \cite{1976PhRvD..13..198P, 1977PhRvD..16.2402P}. The ratio is
\begin{eqnarray}
\frac{L_{\rm H,vis}}{\dot m_{\rm Bondi} u_{\rm init}}
&=& \frac{C_{\rm H}f_{\rm vis} c^2\mpbh^{-2}}{4\pi \lambda \gamma(\gamma-1) G^2 \mpbh^2 \rho /c_s}
\nonumber \\
&=& 4.4\times 10^{-4} \left( \frac{f_{\rm vis}}{0.69} \right) \left( \frac{10^{-16}\,M_\odot}{\mpbh} \right)^{4} \left( \frac{152\,{\rm g\,cm}^{-3}}\rho \right) \left( \frac{c_s}{505\,{\rm km\,s}^{-1}} \right),
~~~~~~~~
\end{eqnarray}
where we substituted $\gamma=\frac53$ and $\lambda=0.25$. We referenced the value of $f_{\rm vis}$ to 0.69, which is the contribution of photons, electrons, and positrons when $\mpbh \ll 4 \times 10^{-17} M_\odot$ (or $k_{\rm B}T_{\rm Hawking}\gtrsim m_ec^2$ \cite{1977PhRvD..16.2402P}). For larger PBH masses, electrons and positron emission is exponentially suppressed, and $f_{\rm vis}$ asymptotes to 0.12 (photons only). We thus expect that in the range of allowed PBH masses ($>3.5\times 10^{-17}M_\odot$) and in stellar cores, Hawking radiation cannot substantially alter the energetics of the flow near the Bondi radius.

\subsection{Constraints from survival of stars}

There are two classes of potential constraints on PBHs arising from capture in stars: constraints arising from stellar survival (observation of a star implies either that it has not swallowed a PBH, or that the swallowed PBH has not yet had time to eat the star); and constraints arising from the directly observable signatures of a star being eaten by a captured PBH (which may take many forms, depending on the nature of the star's final demise).

Let us first consider the stellar survival constraints. Typical conditions in the Milky Way disc have $\rho_{\rm DM}c^2 \sim 0.4\,$GeV cm$^{-3}$ and $\sigma\sim 200$ km s$^{-1}$, so even if PBHs comprise all of the dark matter it is clear that the number of captured PBHs in the lifetime of the Galaxy for a star of a few stellar masses is small ($N$ is of order $10^{-7}$). Even if one hypothesizes a dark matter ``spike'' in the center of the Galaxy, with much higher densities ($\sim 10^4$ GeV cm$^{-3}$, if one can extrapolate $\rho\propto r^{-1}$ from the Solar System in to several tenths of a parsec), the mean number of PBHs captured by a star is still $\sim 10^{-3}$. Therefore stellar survival is not a consideration in these environments.

Globular clusters have been recognized as a potential site for stellar survival constraints, due to their low typical velocity dispersions $\sigma \approx 7/\sqrt3 \approx 4$ km s$^{-1}$ \cite{2013PhRvD..87b3507C, 2013PhRvD..87l3524C}. For a typical neutron star mass of 1.4 $M_\odot$, the bound of Eq.~(\ref{eq:NdotB}) gives an expected number of PBHs captured of unity in 10 Gyr if the PBH density $\rho_{\rm PBH}c^2$ exceeds 30 GeV cm$^{-3}$. Above this density, there is a possible constraint of $f_{\rm PBH}<($30 GeV cm$^{-3}$)/$(\rho_{\rm DM}c^2)$, {\em if} the phase space bound is saturated and {\em if} a captured PBH consumes the neutron star. This was the bound derived in Ref.~\cite{2013PhRvD..87l3524C} in the range of $\mpbh\sim 10^{18}-10^{23}$ g (for lower masses, the PBHs do not sink to the NS center; for higher masses, capture can occur from trajectories that are unbound to the globular cluster and hence are not occupied by PBHs, hence $f\approx 0$ in the integrand of Eq.~\ref{eq:Ntemp}, and the integral is much less than the bound obtained using $f_{\rm max}$). However, as has been noted (e.g.\ \cite{2016PhRvD..94h3504C}), these bounds depend on the globular clusters being formed in early dark matter halos and hence containing substantial non-baryonic dark matter today -- a scenario that is possible but by no means established. Until the question of the origin of globular clusters is resolved, the robustness of PBH constraints based on survival of stars in globular clusters will remain uncertain.

Dwarf galaxies offer a potential constraint, since their velocity dispersions are also low, but they are known to be dark matter dominated. We may make an order-of-magnitude estimate of the maximum capture rate by replacing $\rho_{\rm PBH}$ in Eq.~(\ref{eq:NdotB}) with $\sim C\sigma^2/2\pi Gr^2$, where $r$ is the radius and $C$ denotes an order-unity correction for deviation from the singular isothermal sphere density profile; this gives
\begin{equation}
N\le 2\sqrt{\frac2\pi} \frac{CGMt}{\sigma r^2} \ln\Lambda =
8.3\times 10^{-4}C \frac{M}{M_\odot} \frac{t}{10\,\rm Gyr} \frac{10\,\rm km\,s^{-1}}{\sigma} \left(\frac{0.5\,\rm kpc}{r}\right)^2 \frac{\ln\Lambda}{30}.
\end{equation}
The parameters are scaled to typical circular velocities and half-light radii of the Milky Way satellites (e.g.\ Fig.\ 10 of Ref.~\cite{2017ARA&A..55..343B}); it is thus clear that only a tiny fraction of stars would have captured a PBH in this environment, even if PBHs make up all of the dark matter.

We thus conclude that the survival of stars over a cosmological timescale does not at present rule out PBHs as the principal component of dark matter in any mass range. A constraint might arise in the future if some globular clusters could be definitively shown to have formed in small dark matter halos.

\subsection{Constraints from signatures of stars being destroyed by PBHs}
\label{ss:runaway}

An alternative route to a constraint is the directly observable signatures of a star being destroyed by a PBH. Bondi accretion has $\dot m_{\rm PBH} \propto m_{\rm PBH}^2$, which means that $\mpbh(t)$ runs away to $\infty$ after a finite time $t_{\rm Bondi}$, but of course in a star of finite initial mass this solution will not hold forever and at some point the star is disrupted. The observable signatures depend in great detail on what happens once the PBH grows big enough to have a substantial feedback effect on the star. We are very interested in these signatures because the capture of PBHs by stars followed by runaway growth by accretion is a robust prediction of asteroid-mass PBH dark matter. It applies to the {\em entire} mass range between the evaporation and microlensing constraints, and is thus a promising route to either detect PBHs or rule out the low mass window entirely. We emphasize that to derive a constraint would require robust modeling of the fate of the star after accretion onto the PBH starts to affect its evolution, sufficient to ensure that the resulting event would be detected. Plausible signatures could include (i) an long-lived ultra-luminous phase such as Eddington-limited accretion onto the central black hole \cite{2009arXiv0901.1093R}; (ii) a transient event accompanying the final destruction of the star; (iii) nucleosynthesis patterns from stellar destruction \cite{2017PhRvL.119f1101F}; and (iv) a remnant black hole with a mass less than the maximum mass for a neutron star \cite{2009arXiv0901.1093R}, which may be isolated or in a binary system. A detailed study of these is beyond the scope of this paper, however we are in a position here to estimate the rate of PBH-mediated stellar destruction events.

We now estimate the expected PBH capture rate in the Milky Way. We should replace $M\rho_{\rm PBH}$ in Eq.~(\ref{eq:NdotB}) with $\int \rho_\star \rho_{\rm DM}\,{\rm d}^3{\boldsymbol r}$, which -- using the default {\sc Trilegal} v1.6 model for the stars \cite{2005A&A...436..895G, 2009A&A...498...95V} and the reference model (BjX, Table 1) of Ref.~\cite{2017JCAP...02..007B} for the dark matter -- is $5.8\times 10^{11}$ $M_\odot\,$GeV$\,c^{-2}\,$cm$^{-3}$. If we take the canonical estimate of $\sigma = v_0/\sqrt2$, with $v_0 = 230$ km s$^{-1}$, then we establish a bound of ${\rm d}N/{\rm d}t \le 20$ Myr$^{-1}$. This is an upper bound; if the stars are on circular orbits at velocity $v_0$ and capture PBHs primarily from low relative velocities, then there should be a suppression factor of ${\rm e}^{-v_0^2/2\sigma^2} = 0.37$. We thus take as our fiducial estimate of the capture rate ${\rm d}N/{\rm d}t = 7.4$ Myr$^{-1}$. In particular, we expect $N_{\rm max} \sim 7\times 10^4$ PBH capture events over the history of the Milky Way. For $\mpbh\gtrsim {\rm few}\times 10^{-16}\,M_\odot$, the Bondi time is fast for most stars and the stellar disruption rate today is roughly equal to the capture rate. At lower $\mpbh$, only the low-mass main sequence stars will be disrupted, and the rate may be slower. For example, 14\%\ of the stellar mass\footnote{Including brown dwarfs.} in the Chabrier IMF \cite{2003PASP..115..763C} is below $0.20M_\odot$ and hence will have $t_{\rm Bondi}<10$ Gyr at $\mpbh=3.5\times 10^{-17}\,M_\odot$ (see \S\ref{ss:postcapture}). A further $\sim 10\%$ of the stellar mass in the Milky Way is in white dwarfs (e.g., \cite{2009JPhCS.172a2004N}), which will also have short $t_{\rm Bondi}$. We thus expect that at this lower mass, the stellar disruption rate is less than the capture rate, but by only a factor of 4--10.

This suggests that destruction of stars by PBHs is 3--4 orders of magnitude less common than supernovae. With a density of $\sim 0.01$ Milky Way-size galaxy per Mpc$^3$, we expect the nearest such event in the past 20 years to have been at a distance $D \sim 50$ Mpc; whether such an event should have been seen depends on its observational signature, which has not been robustly modeled to our knowledge.

\section{White dwarf survival}
\label{sec:explosion}

Here we will treat the problem of PBH transit through a carbon/oxygen white dwarf. We will answer two questions: can the passage of a PBH cause enough localized heating by dynamical friction to ignite the carbon; and, can this ignition provoke a runaway explosion.
We expand on the work of Ref.~\cite{2015PhRvD..92f3007G} with the next logical step, i.e., using 1+1D (cylindrical radius + time) hydrodynamic simulations to follow the shock heating during the passage of the PBH. We split this problem into three pieces. First we compute the density profile of the WD and the velocity profile of the incoming PBH. Next, we use a 1+1D Lagrangian hydrodynamic code to study the possible shocks and track the evolution of thermodynamic quantities in the neighborhood of the PBH trajectory, where the initial conditions for the simulations are set by the profiles obtained in the first step. Finally, we follow the arguments in Ref.~\cite{1992ApJ...396..649T} to find out if the shock meets the conditions to start a runaway thermonuclear explosion, e.g., by comparing the specific nuclear energy generation rate to the rate of energy loss due to conduction, and the burning timescale to the Kelvin-Helmholtz instability timescale. (Note that it is not clear if the presence of instabilities will definitively destroy the flame propagation; this would require a full 3+1D simulation and is beyond the scope of this paper.)

As pointed out in Ref.~\cite{2015PhRvD..92f3007G}, the survival of a WD provides two different ways to constrain PBHs. One could put constraints on the abundance of PBHs simply by the fact that we observe WD in a certain mass range. Furthermore, since the passage of PBH could explode WD with masses lower than the Chandrasekhar mass, one could look at the rate of Type Ia supernova to place constraints. Here we will consider only the first type of constraint, since it is not clear whether a successful ignition of the type considered here would lead to a recognizable Type Ia supernova.

\subsection{Velocity and density profiles}
\label{ssec:stage1}

To determine the velocity and density profile of a white dwarf star, we numerically integrated the equation of state assuming zero temperature (since electron degeneracy pressure dominates) \cite{1986bhwd.book.....S}.  The composition a typical C/O white dwarf consists of free electrons with by carbon and oxygen ions; we assume here a 50:50 mixture. Neglecting ion-ion interactions, once can average the atomic number of both atoms, i.e. $\bar{Z} = 7$. The ion density and pressure can be represented by the Fermi parameter $x = p_{F}/m_{e}c$:
\begin{equation}
\rho  =  \frac{n_{e}m_{B}}{Y_{e}} = \frac{m_{B}}{3\pi^{2}\lambda_{e}^{3}Y_{e}}x^{3} 
~~{\rm and}~~
P=\frac{2}{3h^{3}} \int_{0}^{p_{F}} \frac{p^{2}c^{2}}{(p^{2}c^{2} + m_e^{2}c^{4})^{1/2}} 4 \pi p^{2} dp  = \frac{m_{e}c^{2}}{\lambda_{e}^{3}} \phi(x) \, \textup{,}
\end{equation}
where $\lambda_{e} = {\hbar}/({m_{e}c})$ is the electron Compton wavelength, and $\phi(x)$ is given by \cite{1986bhwd.book.....S} 
\begin{equation}
\phi(x) = \frac{1}{8\pi^2} \left[x(1+x^{2})^{1/2}\left(\frac{2x^2}{3}-1\right) + \ln(x + (1+x^2)^{1/2})\right] \, \textup{.}
\end{equation}
We also applied electrostatic corrections to the pressure and density in our calculation. At zero temperature, the ions form a lattice that maximizes the inter-ion separation. Each ion is surrounded by a uniform distribution of electrons such that each spherical shell, known as \textit{Wigner-Seitz cell}, is neutral \cite{1986bhwd.book.....S}. Due to the electron-electron and electron-ion interactions in each of the cells, the pressure and density are reduced by a few percent with respect to the case where electrostatic corrections are ignored. Additionally, we also applied a correction term due to deviations of the electron distribution from uniformity. This provides a decrease towards the density and pressure, but by a significantly smaller contribution \cite{1961ApJ...134..669S}. 
To obtain the density and velocity profiles of the star, we use the metric for a spherically symmetric star:
\begin{equation}
\label{eq:metric}
ds^{2} = - e^{2\Phi(r)}c^{2}dt^{2} + \left[1 - \frac{2Gm(r)}{rc^{2}}\right]^{-1}dr^{2} + r^{2}d\theta^{2} + r^{2}\sin^{2}\theta d\phi^{2} \, \textup{,}
\end{equation}
where $\Phi(r)$ is some scalar function that acts as the ``gravitational potential'' in the classical limit and $m(r)$ is the mass up to some radius. We numerically solve the Tolman-Oppenheimer-Volkoff (TOV) equations using the aforementioned equation of state.

We ultimately need the rate of collisions between PBHs and white dwarfs. To determine this, we need the escape velocity at any shell, and the maximum specific angular momentum $\tilde l$ that a PBH can have and still reach that shell. We restrict ourselves to $m_{\rm PBH}\ll M_{\rm WD}$; it follows that the length scale of the accretion flow and associated shock structure around the PBH as it passes through the white dwarf ($R_A = 2G\mpbh/v^2\sim G\mpbh/v_{\rm esc}^2$) are much smaller than the radius of the white dwarf itself ($\sim GM_{\rm WD}/v_{\rm esc}^2$). It also follows that we can treat the PBH as a test particle in the TOV spacetime generated by the white dwarf.
%We restrict ourselves to $m_{\rm PBH} < 10^{-10}$ $M_{\odot}$ due to previous microlensing constraints. These PBHs will have a Schwarzchild radius of order $R_{s} = 2 m_{\rm PBH}G/c^{2}$ < 295.3 nm. Due to their small size, we can ignore the contributions of the PBHs towards the metric of the white dwarf. 

Assuming we are in the rest frame of a fluid mass element of the white dwarf along the equatorial plane ($\theta = \pi/2$), we can write the equation of motion for a PBH starting at rest at infinity and radially falling through the interior of the white dwarf. By solving for the maximum radial velocity, we arrive at the formula for the escape velocity:
\begin{equation}
v_{\rm esc} = c \sqrt{1 - e^{2 \Phi (r_{\rm min})} } \, \textup{.}
\end{equation}
For reasonable parameters, the initial velocity of an incoming DM particle is small compared to $v_{\rm esc}$, so in the interior of the WD we may take the actual velocity to be the escape velocity, $v\approx v_{\rm esc}$.
We can also compute the Mach number $\mathcal{M} = v_{\rm esc}/c_{s}$ where $c_{s}$ is the speed of sound through the interior of the white dwarf.
The specific angular momentum is then
\begin{equation}
\tilde{l} = \frac{r_{\rm min}v(r_{\rm min})}{\sqrt{1 - v^{2}(r_{\rm min})/c^2}} \, \textup{.}
\end{equation}
For our range of white dwarfs ranging from $0.75 \, M_{\odot} - 1.385 \, M_{\odot}$, we calculated the density profile, and thus, the escape velocity, Mach number, and angular momentum per unit mass. In \S\S\ref{ssec:stage2} and \ref{ssec:stage3}, we will calculate the minimum PBH mass for ignition and the rate of collisions assuming the PBH will pass through the white dwarf, reach a minimum radius, and then leave the star.

\subsection{Thermal effects on WD materials by a passing PBH}
\label{ssec:stage2}

Given the kinematics of PBHs passing through WDs, we are now able to calculate the thermal effects generated on small scales, which will be crucial to determining whether part of the WD can be shocked and significantly heated up, and consequently, whether a runaway thermonuclear explosion can occur.

\begin{figure}
\centering
\includegraphics[width=\textwidth]{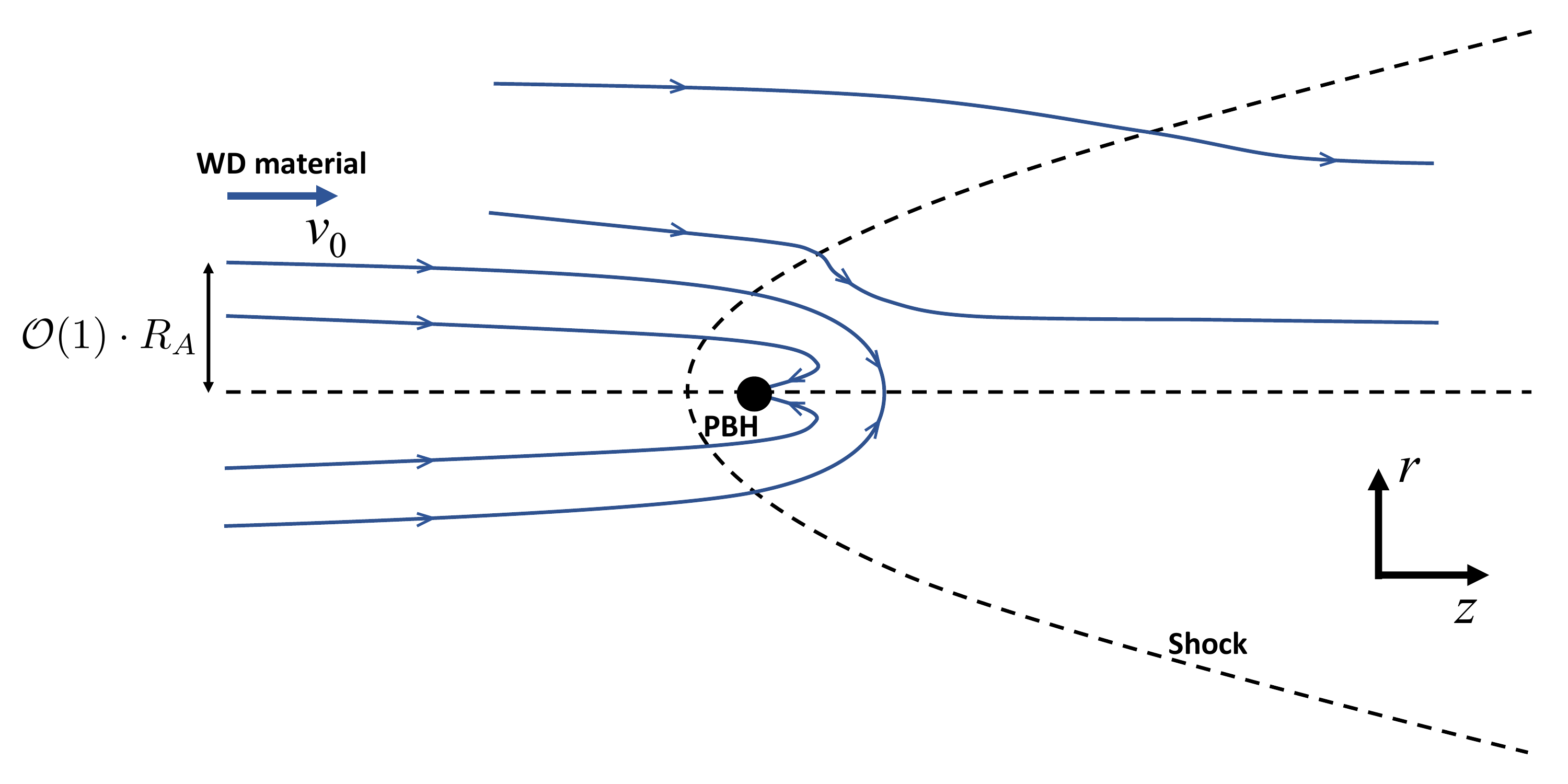}
\caption{A schematic diagram for the PBH passing through the WD materials, creating a shock. The materials with impact parameter less than about the accretion radius eventually get eaten by the PBH, while materials farther away get shocked, compressed and heated. The diagram is shown in the rest frame of the PBH.}
\label{fig:shock}
\end{figure}

We set up a local 1+1D Lagrangian hydro simulation \citep[see \eg,][for a review]{1992CMAME..99..235B} to capture the evolution of thermal properties of the fluid around the PBH trajectory. In the rest frame of the PBH, the fluid is initially steadily flowing towards the $+z$ direction with velocity $v$. In the simulation, we scale quantities by defining $c_{s,\infty} = \rho_{\infty}= R_c = 1$, where $c_{s}$ is the sound speed, $\rho$ is the density of the WD. We use subscript $\infty$ to denote the initial background values, which are equivalent to the values at infinite impact parameter $r=\infty$. $R_c = 2Gm_{\rm PBH}/(c_{s,\infty} v)$ is the critical radius where the inward radial velocity of the fluid parcel due to PBH's gravity is equal to the sound speed. Another important scale is the accretion radius $R_A = 2Gm_{\rm PBH}/v^2 = R_c/\mathcal{M}$. Fluid parcels with impact parameters $b$ smaller than $R_A$ will be strongly deflected from their trajectory. In fact, fluid parcels with $b$ less than a critical value $b_c$ will be eaten by the PBH, thus irrelevant to us. $b_c$ is given by
\begin{equation}
b_c = \sqrt{\frac{\dot{m}}{\pi \rho_\infty v}} = \sqrt{\alpha} R_A~,
\end{equation}
where $\dot{m} = 4\pi\alpha(Gm_{\rm PBH})^2\rho_{\infty}/v^3$ is the mass accretion rate, and the parameter $\alpha$ is approaching 1 at $\mathcal{M}\gg 1$ limit \citep{1944MNRAS.104..273B,1971MNRAS.154..141H}\footnote{Note that our definition of $\alpha$ is consistent with \cite{1971MNRAS.154..141H} and is half of that defined in \cite{1944MNRAS.104..273B}.}. We will only care about the fluid outside of the accretion radius $R_A$, since material at smaller impact parameter falls into the black hole and it does not matter whether it burns. See Figure \ref{fig:shock} for the physical picture.

Computations could in principle be done in the frame of the PBH (so that we work with radius $r$ and downstream position $z$), or alternatively we could work in the initial frame of the material so that time since passage $t = z/v$ is the independent variable instead of $z$. In the high Mach number limit, it is more convenient to take the latter perspective; except in the vicinity of the PBH (near $R_A$), the flow will everywhere be supersonic, the partial differential equations should be hyperbolic, and the time evolution picture is appropriate.

Our code tracks the evolution of 1200 concentric mass shells following passage of the PBH at $t=0$; by default, it uses time steps $\Delta t = 4\times 10^{-3}$ (in the code unit $t_{\rm code} = R_c/c_s$) and runs up to 5000 steps. The code tracks the Eulerian radius $r$ of each shell, its radial velocity $v_r=dr/dt$, and the specific energy $\epsilon_s\equiv \epsilon + v_r^2/2$. The evolution in each time step is calculated with adaptive step sizes $\delta t = \frac{1}{40}\min\lbrace\vert\frac{dr}{dv}\vert, \frac{r}{v_{\rm sp}}\rbrace$, where $v_{\rm sp}=\vert v \vert+\sqrt{\frac{5}{3}(\frac{5}{3}-1)\epsilon_s}$ is the spread speed. We implement artificial viscosity by subtracting a smooth term, $m \, \Psi \Delta v$, from the momentum flux, where $\Delta v$ is the difference in velocity of the adjacent cells, and $\Psi$ is a function that determines how much damping is needed at each boundary by computing the finite velocity difference. The mass shells are chosen in a way that the impact parameters are logarithmically spaced and the outer radius of the 400th shell has initial radius $R_c$. Initially, the WD temperature $T$ is much lower than the Fermi temperature $T_F$, so the fluid is treated as a relativistic degenerate Fermi gas with zero temperature (\ie $\gamma = 4/3$ polytrope). The initial pressure is then $P_{\rm init}=c_s^2\rho_{\infty}/\gamma$, and the specific internal energy is given by $\epsilon_{\rm in,init} = P_{\rm init}/[\rho_\infty(\gamma-1)]$. Initially we set each shell's Eulerian position to equal its Lagrangian position (so the density is 1), the specific internal energy is $\frac94$ (chosen to agree with $c_{s,\infty}=1$), and the initial radial velocity is $v_r=-1/r$ (the result computed in the impulse approximation).

During the evolution, the fluid may be compressed by the PBH, and $T$ may rise non-negligibly comparing to $T_F$, where the constant polytropic equation of state no longer holds and the heat capacity of ions may also be important. To run a Lagrangian hydrodynamic code, it is necessary to derive an equation of state function $P(\rho,\epsilon)$. Our implementation of the equation of state is described in Appendix~\ref{app:code_unit}, and is facilitated by use of the auxiliary variable $J = \mu_e/kT$ (which goes to $+\infty$ for perfectly degenerate material at zero temperature, and 0 in the limit that the leptons become a pair plasma). The equation of state includes ion thermal pressure (the ideal gas law), as well as relativistic electrons and positrons as appropriate for $x\gg 1$. The current implementation does not yet include radiation pressure.

\begin{figure}
\centering
\vspace{-35pt}
\includegraphics[width=\textwidth]{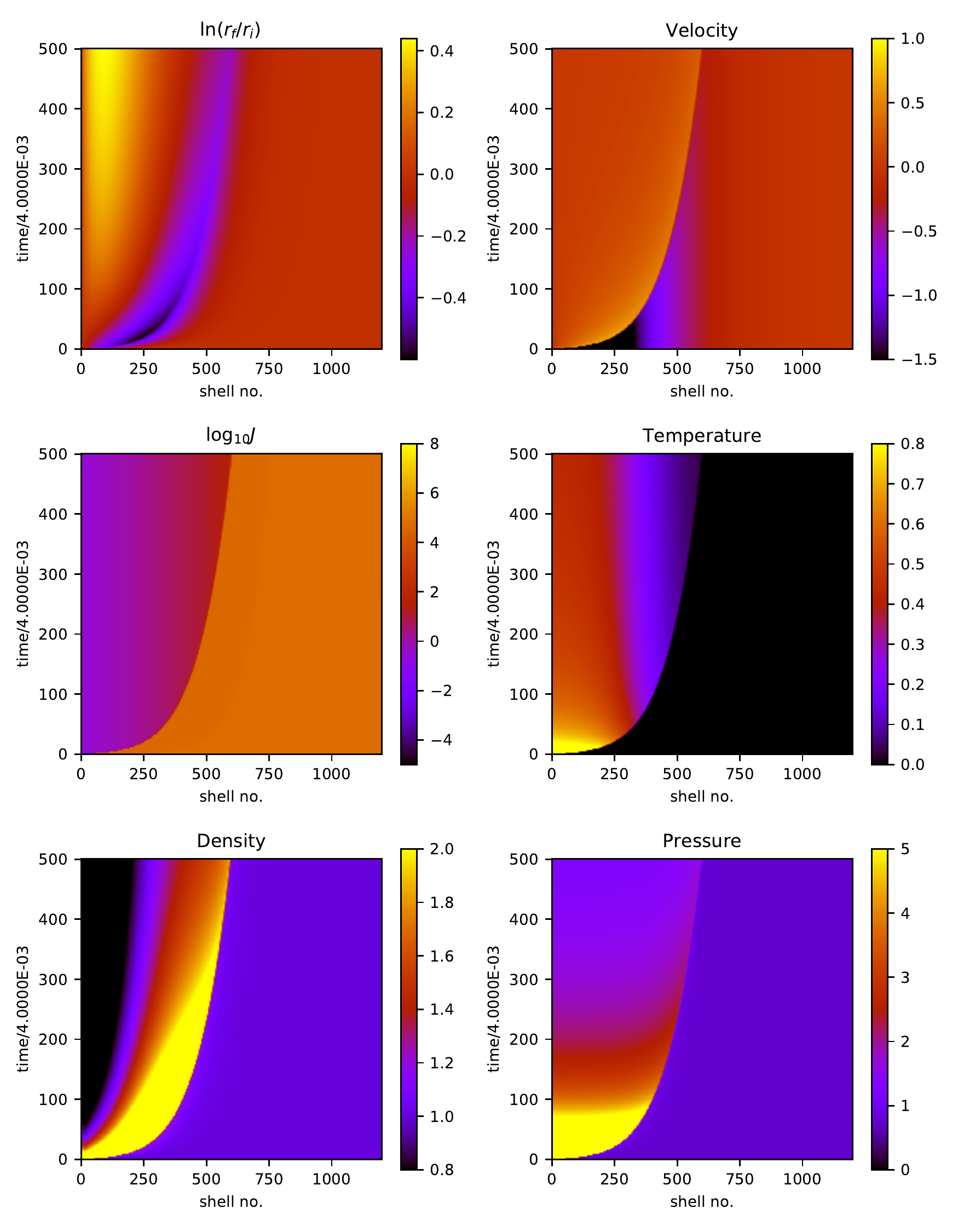}
\vspace{-28pt}
\caption{The simulated evolution of the logarithm ratio of the positions (i.e., Eulerian divided by Lagrangian radius coordinate), radial velocity, thermodynamic parameter $J$, temperature, density and pressure of the 1200 mass shells. The critical radius is at shell 400. We only show the first 500 of the 5000 time steps we run, to illustrate the micro-scale thermal evolution of mass shells during the PBH passage, where each time step is $4\times 10^{-3}R_c/c_s$. The impact gets weaker with time and in mass shells farther out. All the quantities are in our code units, where $c_{s,\infty}=\rho_\infty=R_c=1$.}
\label{fig:simu1}
\end{figure}

\begin{figure}
\centering
\includegraphics[width=0.8\linewidth]{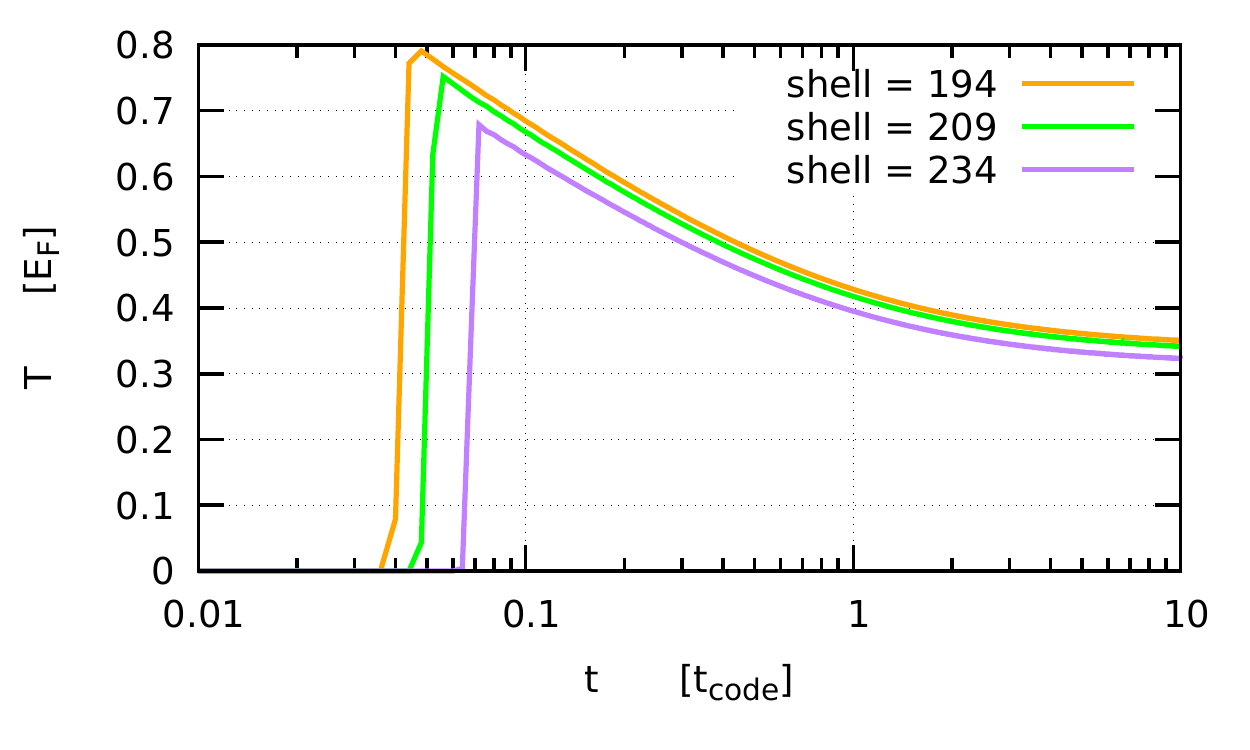}
\caption{Temperature time evolution for three different shells. The shells are chosen so that they correspond to the accretion radius for three different mach numbers: $\mathcal{M} = 2.60$ (purple), $3.00$ (green) and $3.24$ (orange).}
\label{fig:temptime}
\end{figure}

\begin{figure}
\centering
\includegraphics[width=0.8\linewidth]{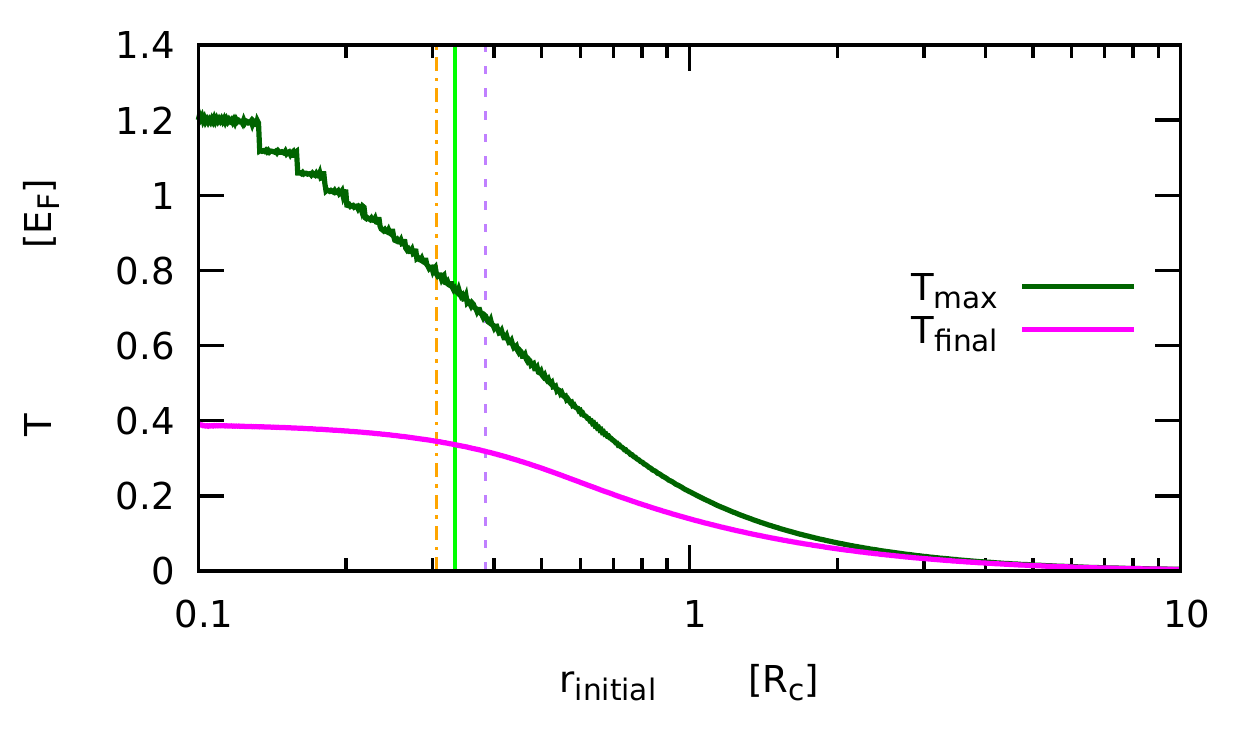}
\caption{Final and maximum temperatures reached by each cell as a function of Lagrangian radius. The initial radius for the shells from Fig.~(\ref{fig:temptime}) have been highlighted preserving the color code.}
\label{fig:tempmaxfinal}
\end{figure}

In Figure \ref{fig:simu1} we show the evolution of the 1200 mass shells. Naturally, the region closer to the PBH reaches higher temperature, followed by a rapid drop as time passes and the material expands radially. Furthermore, one must ignore shells inside the accretion radius since they will get ``eaten''. We find that the maximum temperature is achieved right at the accretion radius, which corresponds to the 209-th shell if $\mathcal{M} = 3$. In Fig.~(\ref{fig:temptime}), we show the temperature evolution of that shell and two additional shells that have $\mathcal{M} = 2.60$ (234-th shell, outer radius $r=0.385R_c$) and $3.24$ (194-th shell, $r=0.305R_c$). The temperature quickly reaches its maximum, then swiftly decreases and approaches an asymptotic final temperature -- in terms of the Fermi energy -- of $\approx 0.4\,\textup{E}_{\rm F}$. Additionally, in Fig.~(\ref{fig:tempmaxfinal}), we present the maximum and final temperature as functions of the mass shells' initial radii.
 
We test convergence of our hydro simulations by running a lower time resolution simulation with double the time stepsize, i.e. $\Delta t = 8 \times 10^{-3}$ and up to $2500$ steps. We find that at each step the fractional differences of the thermodynamic quantity $J$, and the temperature are within $10^{-3}$, and within $10^{-4}$ for the pressure. 

\subsection{Ignition and runaway explosion?}
\label{ssec:stage3}

The destruction of a WD will occur only if the energy generation due to carbon fusion in the downstream material exceeds the losses due to conduction. In the case of burning in the ``wake'' of a PBH passage, there may be an additional requirement that the energy injection timescale due to burning be faster than hydrodynamic instabilities that mix the material with cooler plasma and thus suppress the energy generation rate.

To determine whether the carbon fusion is ignited, we focus on the major nuclear reactions, \ie,
\begin{equation}
^{12}{\rm C}\,+\,^{12}{\rm C} \rightarrow \left\lbrace\begin{array}{lclcl}
^{20}{\rm Ne}&\,+&\,\alpha &\,+&\,4.621\,{\rm MeV}\\
^{23}{\rm Na}&\,+&\,p &\,+&\,2.242\,{\rm MeV} \\
^{23}{\rm Mg}&\,+&\,n &\,-&\,2.598\,{\rm MeV} \\
\end{array}\right.\;,
\end{equation}
which have yields 0.56/0.44/0.00 for $T_9<1.75$, 0.50/0.45/0.05 for $1.75<T_9<3.3$, and 0.53/0.40/0.07 for $3.3<T_9<6.0$, where $T_9\equiv T/10^9\,$K \citep{1988ADNDT..40..283C}. We further assume no significant change in the branching ratio above $T_9=6.0$. These considerations allow us to calculate the mean energy release for each temperature range as $\bar{Q}=3.574,3.190,3.164\,$MeV per reaction, respectively. The reaction rate is given by \cite{1988ADNDT..40..283C} as
\begin{equation}
\lambda(T_9) = 4.27\times 10^{26}\frac{T_{9a}^{5/6}}{T_9^{3/2}} \exp\left(-\frac{84.165}{T_{9a}^{1/3}}-2.12\times 10^{-3} T_9^3\right)\,{\rm cm}^3\,{\rm mol}^{-1}\,{\rm s}^{-1}~,
\end{equation}
for $0.001<T_9<10$, where $T_{9a}\equiv T_9/(1+0.0396 T_9)$, and we assume the formula holds for higher temperatures.
The specific nuclear energy generation rate $\dot{\epsilon}_{\rm nuc}$ is then given by $\dot{\epsilon}_{\rm nuc} = \rho Y_{12}^2\bar{Q}\lambda$,
where $Y_{12}=1/28$ is the mole number of $^{12}$C per gram of the material. Written in c.g.s.\ units, we have
\begin{equation}
\label{eq:nucrate}
\dot{\epsilon}_{\rm nuc} = \rho Y_{12}^2\frac{\bar{Q}}{\rm MeV}\frac{\rm MeV}{\rm erg}\frac{N_A}{2}\lambda~~{\rm erg}\,{\rm g}^{-1}\,{\rm s}^{-1}~.
\end{equation}
For temperature $T_9= 5.59$, density $\rho= 2.57 \times 10^7 \, \textup{g cm}^{-3}$, we have $\dot{\epsilon}_{\rm nuc}= 3.83 \times 10^{26} \, {\rm erg}/{\rm g}/{\rm s}$.

Adding the screening effect of the electrons can enhance the nuclear reaction rate (Eq.~\ref{eq:nucrate}) due to less repulsion, hence smaller distances, between nearby nuclei. To estimate this enhancement, we reproduce the procedure present in the alpha-chain reaction networks\footnote{We use the 19 isotope chain, \url{http://cococubed.asu.edu/code_pages/burn_helium.shtml}, with the code described in \cite{1978ApJ...225.1021W}.}. For our purposes we are interested in the regime of degenerate screening, so we adopt the method for strong screening, which is appropriate when density is high. The method and physics are described in \cite{1977ApJ...218..477I,1977JSP....17..357J,1978ApJ...226.1034A,1979ApJ...234.1079I}. Taking this screening enhancement into account, and for typical values of $T_9 = 5.59$ and $\rho = 2.57 \times 10^7 \, \textup{g} \,\textup{cm}^{-3}$, we obtain  
\begin{equation}
\label{eq:nucratescreen}
\dot{\epsilon}_{\rm nuc} |_{\rm screen} = \dot{\epsilon}_{\rm nuc} e^{H_{12}} = 5.07 \times 10^{26} \ \textup{erg} \, \textup{g}^{-1} \, \textup{s}^{-1} \, \textup{,}
\end{equation}
where the screening function evaluates to $H_{12} = 0.28$ for these values.

Ignition occurs when the energy generation rate is higher than the energy loss rate, which is dominated by thermal conduction for white dwarf interiors \citep{1996stre.book.....K}.  We use the electron thermal conductivity of Ref.~\cite{1980SvA....24..303Y} (see discussion in Ref.~\cite{1992ApJ...396..649T}). For a mass shell with Lagrangian radius $r$, the specific conductive energy loss rate is given by
\begin{equation}
\dot{\epsilon}_{\rm cond} = \frac{2D_{\rm cond}T}{\rho r^2}~,
\end{equation}
where $D_{\rm cond}$ is the thermal conductivity, predominantly contributed by the electron-ion scattering, \ie, $D_{\rm cond} = \pi^2k_{B}^2 T n_{e^-}/(3m_{*} \nu_{\rm ei})$, in which the effective electron mass $m_{*}=m_e\sqrt{1+x^2}$, the electron-ion collision frequency $\nu_{\rm ei}$ and the Coulomb integral $\Lambda_{\rm ei}$ are determined by
\begin{equation}
\nu_{\rm ei} = \frac{4\alpha^2 m_e c^2\bar{Z}\Lambda_{\rm ei}}{3\pi\hbar}(1+x^2)^{1/2}~,~~\Lambda_{\rm ei} = \ln\left[\left(\frac{2}{3}\pi\bar{Z}\right)^{1/3}\left(\frac{3}{2}+\frac{3}{\Gamma_{\rm e}}\right)^{1/2} - \frac{x^2}{2(1+x^2)}\right]~,
\end{equation}
respectively, where $\Gamma_{\rm e}=\alpha\hbar c(4\pi n_{e^-}/3)^{1/3}/(k_B T)$ is the dimensionless plasma coupling parameter and $\alpha=1/137$ is the fine-structure constant. Therefore, we can obtain the specific conductive energy loss rate as a function of density and temperature of the WD material.

We must also compare the burning timescale to the hydrodynamic instability timescale to determine whether a runaway explosion might occur. The burning timescale is estimated as
\begin{equation}
\tau_{\rm burn} = \frac{c_p T}{\dot{\epsilon}_{\rm nuc}}= \frac{(c_{p,{\rm ion}}+c_{p,e^{-}}) T}{\dot{\epsilon}_{\rm nuc}}~,
\end{equation}
where $c_p$ is the specific heat capacity at constant pressure which is mainly contributed by ions and partially degenerate electrons. The ion part is well-described by the ideal gas, \ie, $c_{p,{\rm ion}}=5\mathcal{R}/2=(2.079\times 10^8/\bar{A})\,{\rm erg}/{\rm g}/{\rm K}~$, where $\mathcal{R}$ is the gas constant. The electron part can be estimated as
\begin{equation}
c_{p,e^{-}}=c_{v,e^{-}}\left[1+\frac{\pi^2}{3}\left(\frac{k_B T}{E_{\scriptscriptstyle\rm F}}\right)^2\right]=\frac{\pi^2}{2}\mathcal{R}\frac{k_B T}{E_{\scriptscriptstyle\rm F}}\left[1+\frac{\pi^2}{3}\left(\frac{k_B T}{E_{\scriptscriptstyle\rm F}}\right)^2\right]~.
\end{equation}
At zero temperature limit, the electrons do not contribute. However, after being shocked, the temperature can get so high that the electron contribution may become dominant.

The hydrodynamic instability in this case arises from the shear between neighboring shells. After the PBH passes by, shells with smaller impact parameters are expected to be ``dragged along'' behind the PBH, i.e., they should have $v_z<0$ in the frame of the WD. The Kelvin-Helmholtz (KH) instability develops on a timescale of $\tau_{\scriptscriptstyle\rm KH}\sim \vert\bm{\nabla}v_z\vert^{-1}$, where the $z$-direction is along the PBH velocity. Fortunately, even though our code is 1+1D, we can estimate $v_z$ using simple physical arguments. In the rest frame of the PBH, and assuming a time-steady flow, a fluid parcel has conserved specific total energy $\mathcal{H}$, \ie
\begin{equation}
\mathcal{H}\equiv u+\frac{1}{2}v^2|_{\rm PBH\,frame}+\Phi+\frac{P}{\rho}~,
\end{equation}
where $u$ is the specific internal energy, $v|_{\rm PBH\,frame}$ is the velocity of the parcel in the PBH frame, and $\Phi$ is the gravitational potential. In our case, $v^2|_{\rm PBH\,frame} = ({\mathcal M}+v_z)^2 + v_r^2$. In terms of the code units, initially we have $u=9/4$, $v=\mathcal{M}$, $\Phi=0$, and $P/\rho = 3/4$. After the PBH has long passed, assuming the parcel has $u_f, P_f, \rho_f$, velocity $\mathcal{M}-v_z$, and using the conservation law, we obtain
\begin{equation}
v_z = -\frac{1}{\mathcal{M}}\left(u_f + \frac{P_f}{\rho_f} -3 \right)~,
\end{equation}
where $v_z\ll \mathcal{M}$, and we have neglected the $v_z^2$ term. We run the simulation up to 5000 time steps when the final velocity gradients have stabilized. The minimal KH instability timescale is achieved near the accretion radius, where $\tau_{\scriptscriptstyle\rm KH}\sim 0.1 \, t_{\rm code} $. For comparison a typical value of the instability timescale is $6.14 \times 10^{-11} \,\textup{s}$ for $c_s=4.23 \times 10^{8} \,\textup{cm s}^{-1}$, $\mathcal{M}=2.58$, and $m_{\rm PBH}=4.53 \times 10^{23} \,\textup{g}$. In contrast, for $T_9 = 5.59$ and $\rho=2.57 \times 10^7 \,\textup{g cm}^{-3}$, the burning timescale is $2.09 \times 10^{-10} \,\textup{s}$. Therefore, for these particular values, convection might be able to destabilize the flame, preventing runaway explosion.

At very high temperatures, possible endothermic reactions may serve as an additional way of halting the ignition. For a WD of mass $ 1.385 \,\textup{M}_\odot$ the shock may heat the local fluid up to $\sim 4.1 \times 10^{10} \,\textup{K}$, while our fuel, $^{12}$C, breaks into $\alpha$ particles with a binding energy of $B = 7.4 \,\textup{MeV}$ \cite{1990NuPhA.506....1A}; above $\sim 300$ keV, dissociation is thermodynamically favored. The tabulated rate coefficient for $^{12}{\rm C} + \gamma (+\gamma + ...) \rightarrow 3\alpha$ \cite{1988ADNDT..40..283C} rises to $10^{10}\,$s$^{-1}$ (so a dissociation time of $10^{-10}\,$s, comparable to the shock passage time of a PBH) at $T=1.4\times 10^{10}\,$K. In order to avoid this issue, and the consequent need to follow the reaction network, we simply mark the region exceeding $T=1.4\times 10^{10}\,$K in Figure~\ref{fig:phase_space} as the ``carbon dissociation'' region. Note that this occurs only for very massive white dwarfs.
%Note that very massive WDs are not common, thus we choose an upper limit of $M_{\rm WD} \leq 1.385 \ \textup{M}_\odot$.

Loss of energy to neutrino cooling may in principle further reduce the chance of WD explosions at very high temperatures and densities. At temperatures of $\gtrsim$ few$\times 10^9$ K and relevant densities of up to $\sim 10^9$ g/cm$^3$, the dominant neutrino cooling mechanism is $e^++e^-\rightarrow \nu+\bar\nu$ (see Fig.~1 of Ref.~\cite{2001PhR...354....1Y}). The cooling time gets shorter at higher temperature, but even in the relativistic non-degenerate limit ($T\gg T_{\rm F}\gg m_ec^2/k_{\rm B}$), the cooling rate is $Q = 3.6\times 10^{24}T_{10}^9$ erg/cm$^3$/s (Eq.~30 of Ref.~\cite{2001PhR...354....1Y}), which -- given an energy density of $\rho u = \frac{11}4a_{\rm rad}T^4 = 2.1\times 10^{26}T_{10}^4$ erg/cm$^3$, including pairs as well as photons in the energy density -- implies a cooling time of $t_{\rm cool} = u/Q = 57T_{\rm 10}^{-5}$ s. Thus we expect neutrino losses to be insignificant on the timescales for flow near a PBH.

\begin{figure}
\centering
\includegraphics[width=0.8\textwidth]{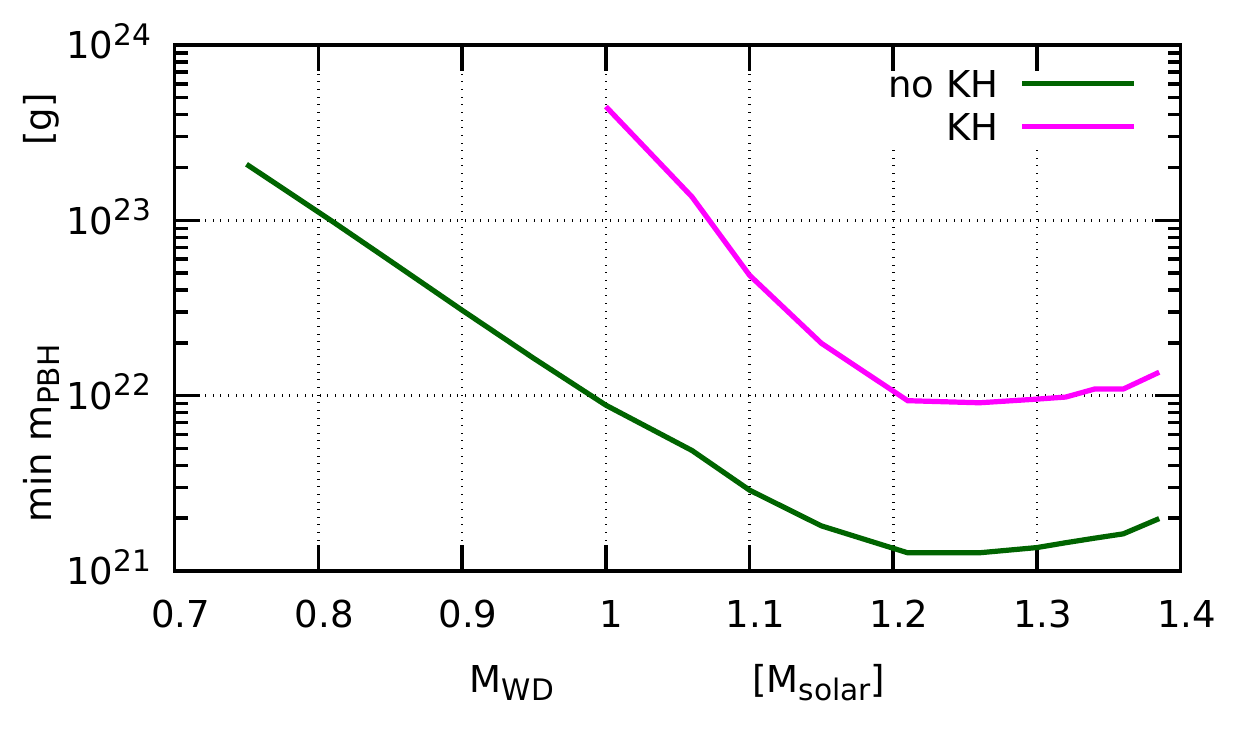}
\caption{The minimum primordial black hole mass needed to achieve thermonuclear runaway for specific WD total mass. For comparison purposes we include the model without considering convection losses by KH instabilities (in green).}
\label{fig:min_pbh}
\end{figure}

\begin{figure}
    \centering
    \includegraphics[width=0.85\linewidth,keepaspectratio]{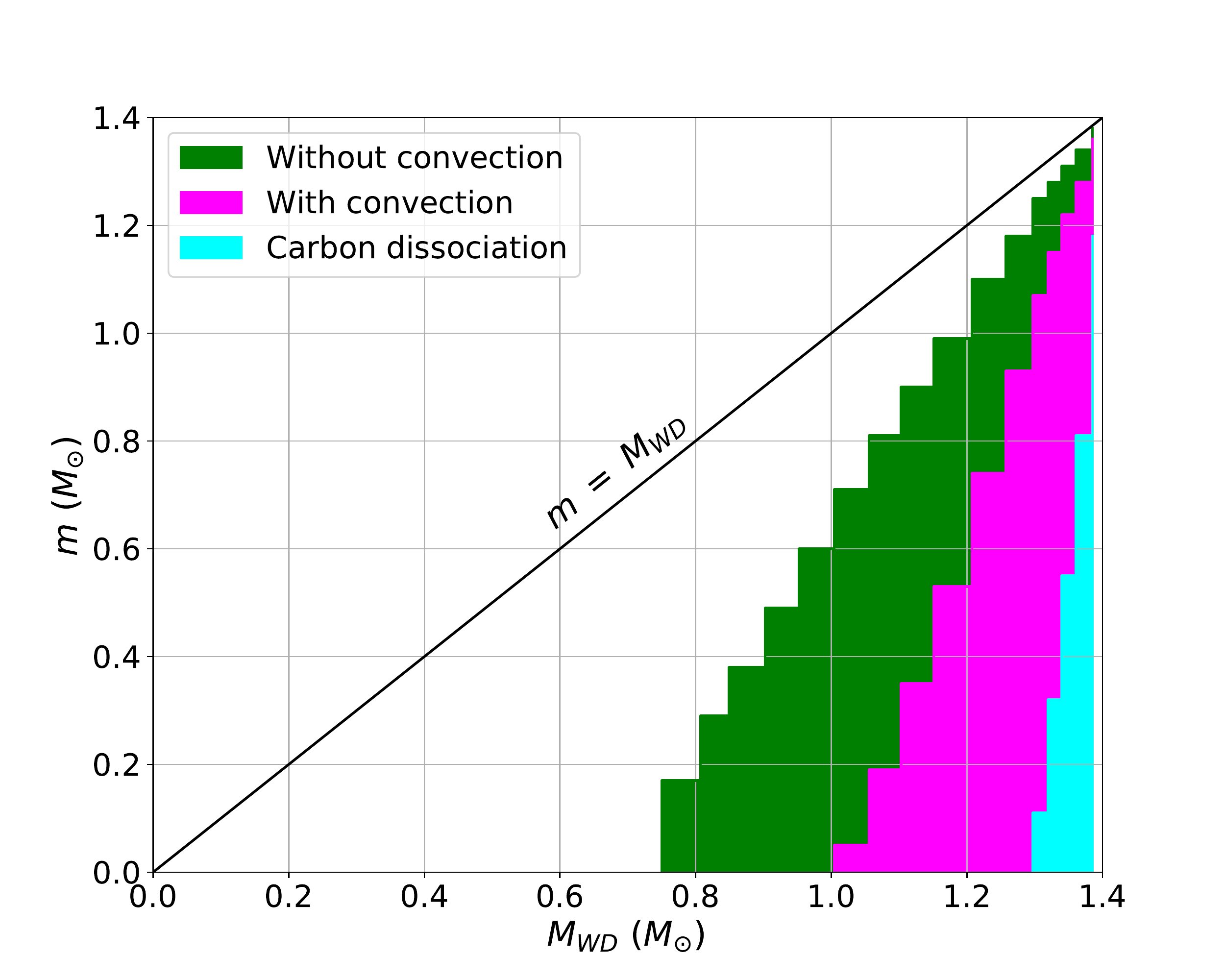}
    \caption{The parameter space where ignition is possible ranging from $0.75 - 1.385 \ M_{\odot}$. Each total WD mass we considered is separated by $\Delta M_{\rm WD} \ \approx \ 0.05 \ M_{\odot}$ while the resolution between each mass shell is $ \Delta m \ \approx \ 0.01 \ M_{\odot}$. The green region is the parameter space where ignition can take place if we ignore the Kelvin-Helmholtz instability but that has $\tau_{\rm burn}>\tau_{\rm KH}$, while the magenta region has $\tau_{\rm burn}<\tau_{\rm KH}$ and hence ignition is robust against this instability. The cyan region is the parameter space where carbon dissocociation into $\alpha$ particles occurs.}
    \label{fig:phase_space}
\end{figure}

In Fig.~\ref{fig:min_pbh}, we plot the minimum required PBH mass for thermal runaway explosion produced by dynamical friction from the passage of a PBH through a WD of a given total mass. The minimum PBH mass is in principle a function of both $M_{\rm WD}$ and the mass shell $m(r)$ where we attempt ignition; in this figure, we take the smallest value of the minimum PBH mass. We show a lower curve (``no KH'') that ignores the Kelvin-Helmholtz instability, and an upper curve (``KH'') that requires the burning time to be shorter than the Kelvin-Helmholtz instability time. Again, the upward-sloping trend at larger masses is due to the inner shells getting to the temperature needed for carbon dissociation. 

In Fig.~\ref{fig:phase_space}, we show the range of parameter space where ignition can occur. We considered a range of white dwarf masses from 0.75--1.385 $M_{\odot}$ with each subsequent WD having a mass larger by $\Delta M_{WD} \ = \ 0.05 \ M_{\odot}$ except that we also considered a 1.32, 1.34, 1.36, and a 1.385 $M_{\odot}$ WD into our analysis. For each WD, we considered ignition in each mass shell where the resolution between mass shells is $\Delta m \ = \ 0.01 \ M_{\odot} $. We considered both requiring and not requiring $\tau_{\rm burn}<\tau_{\rm KH}$; including this criterion significantly restricts the parameter space of where ignition occurs. In regions where $T \geq 1.4\times10^{10}  \ \rm{K}$, carbon dissociation occurs; we are not able to determine whether ignition occurs in this case, but a careful analysis of this parameter space would be of great interest in the future. 

\subsection{Ignition rate and PBH constraints}

\begin{figure}
\begin{multicols}{2}
    \centering
    \includegraphics[width=\columnwidth]{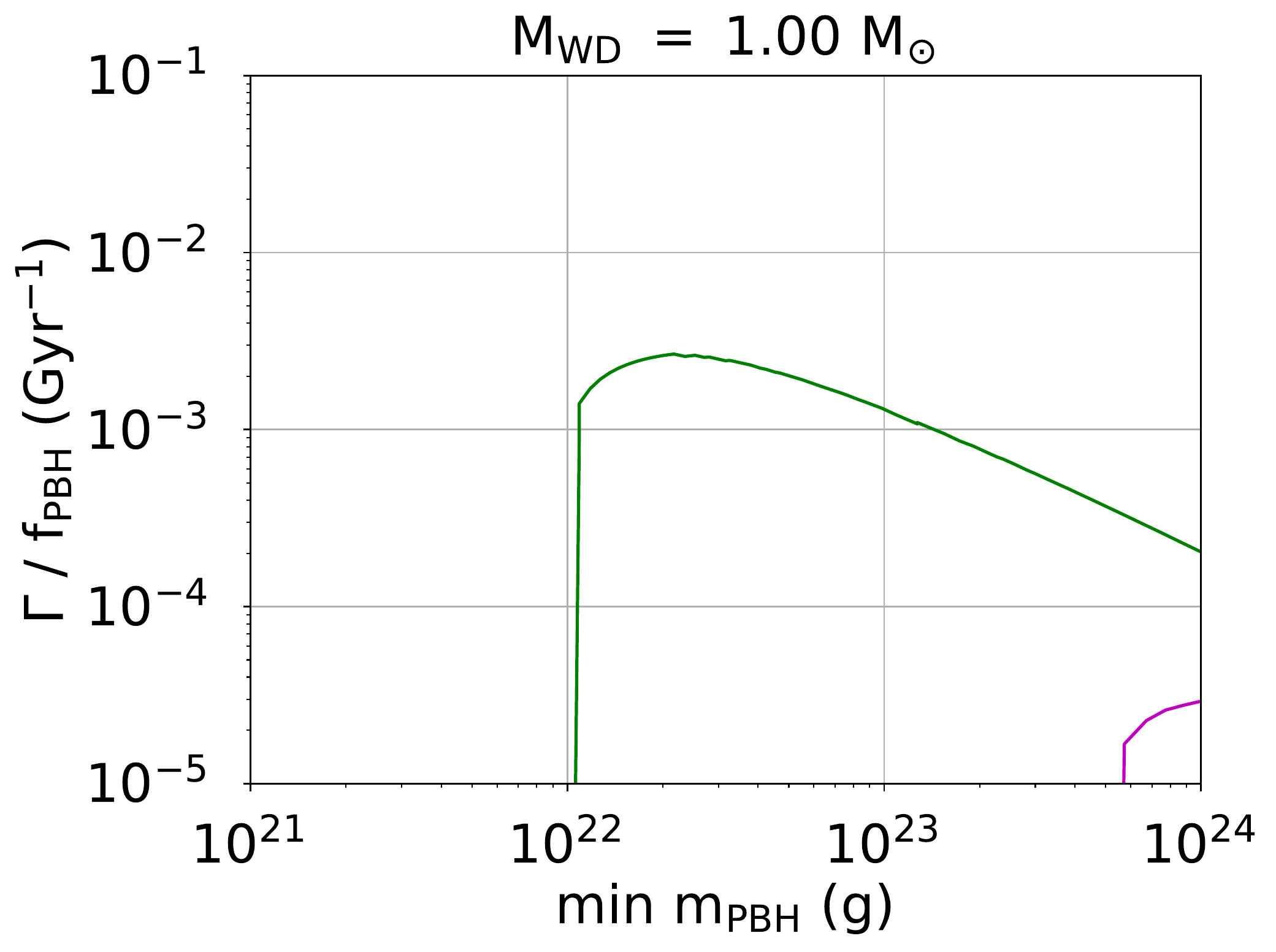}
    \includegraphics[width=\columnwidth]{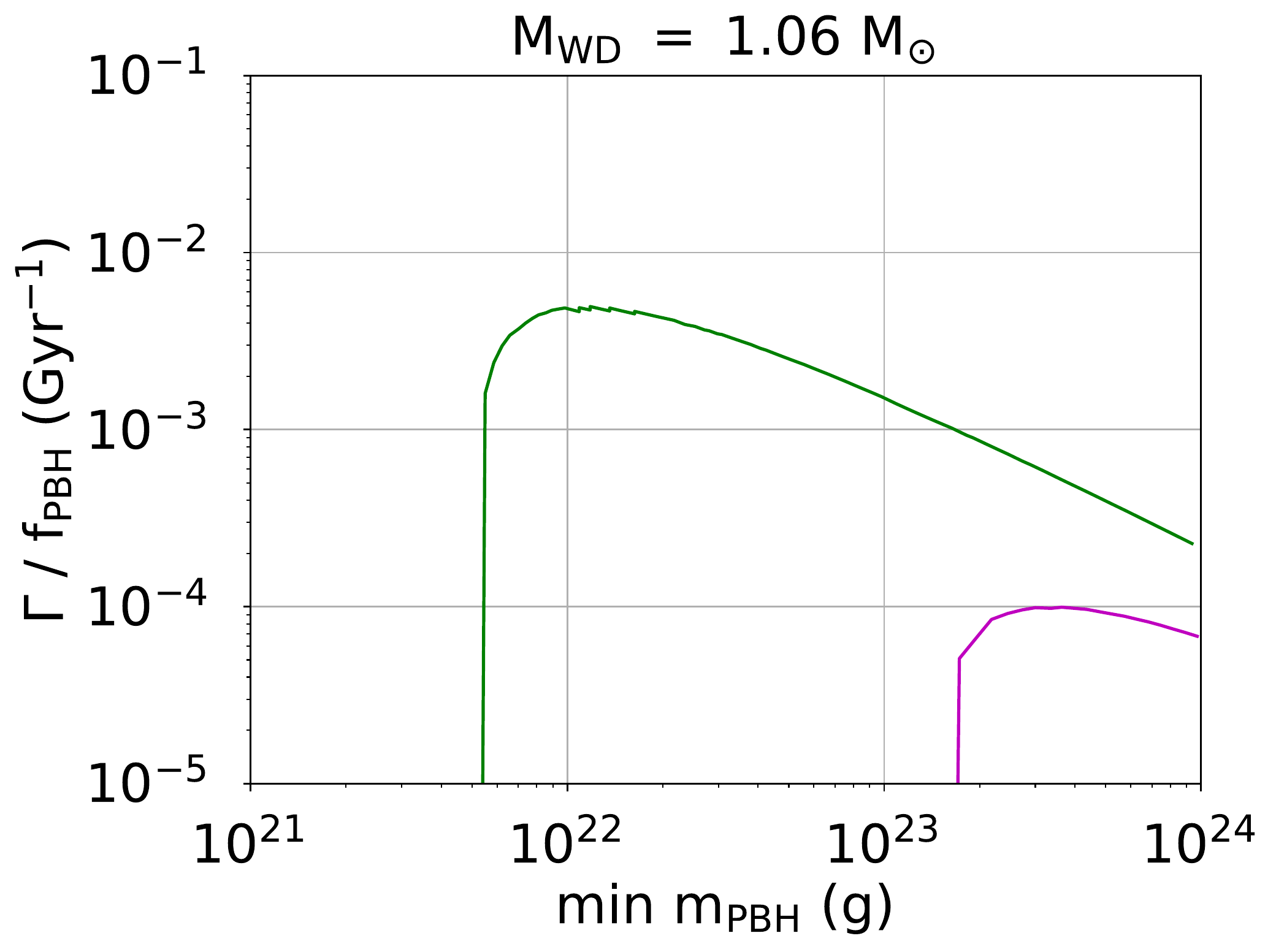}
\end{multicols}

\begin{multicols}{2}
    \centering
    \includegraphics[width=\columnwidth]{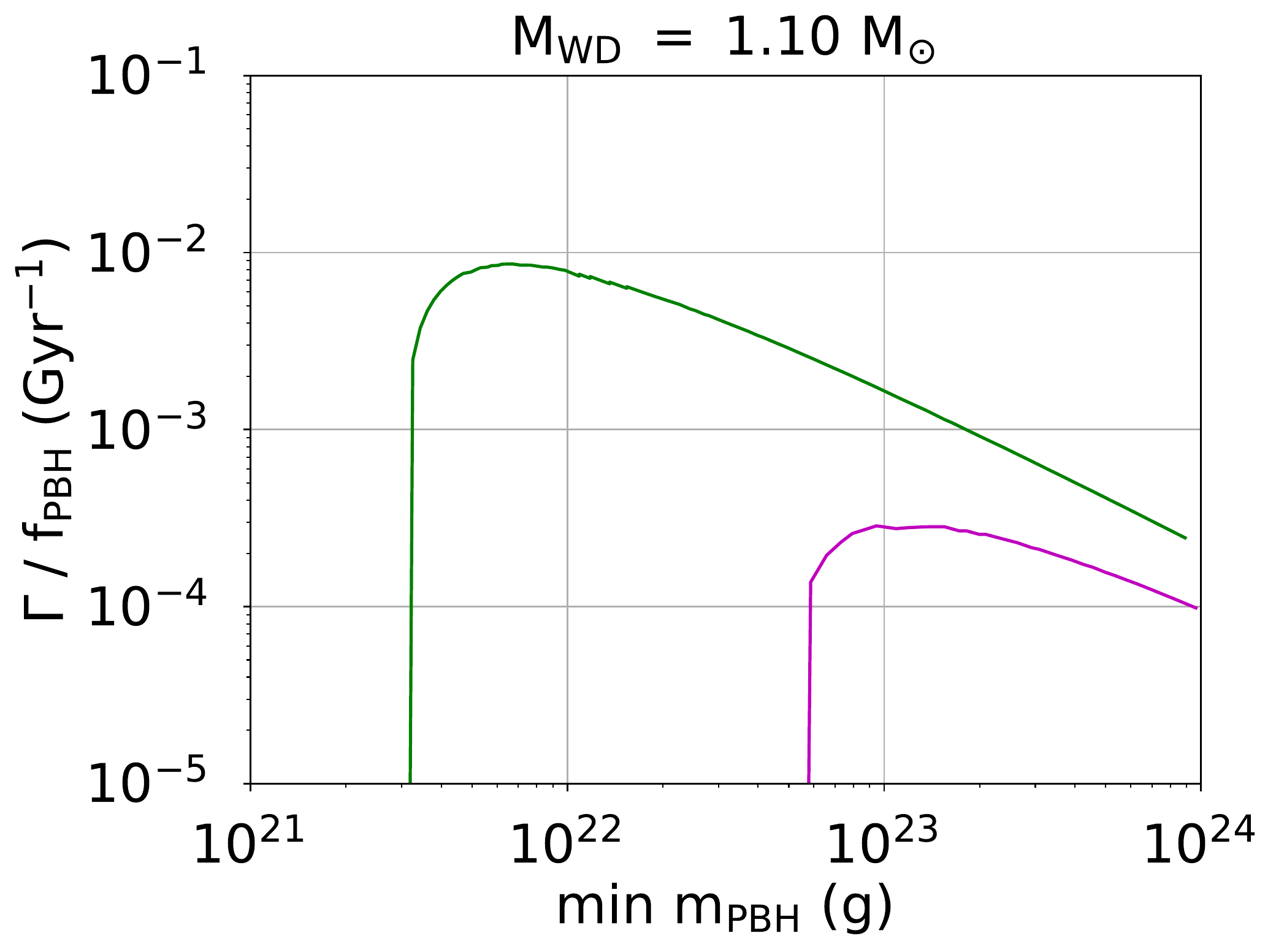}   
    \includegraphics[width=\columnwidth]{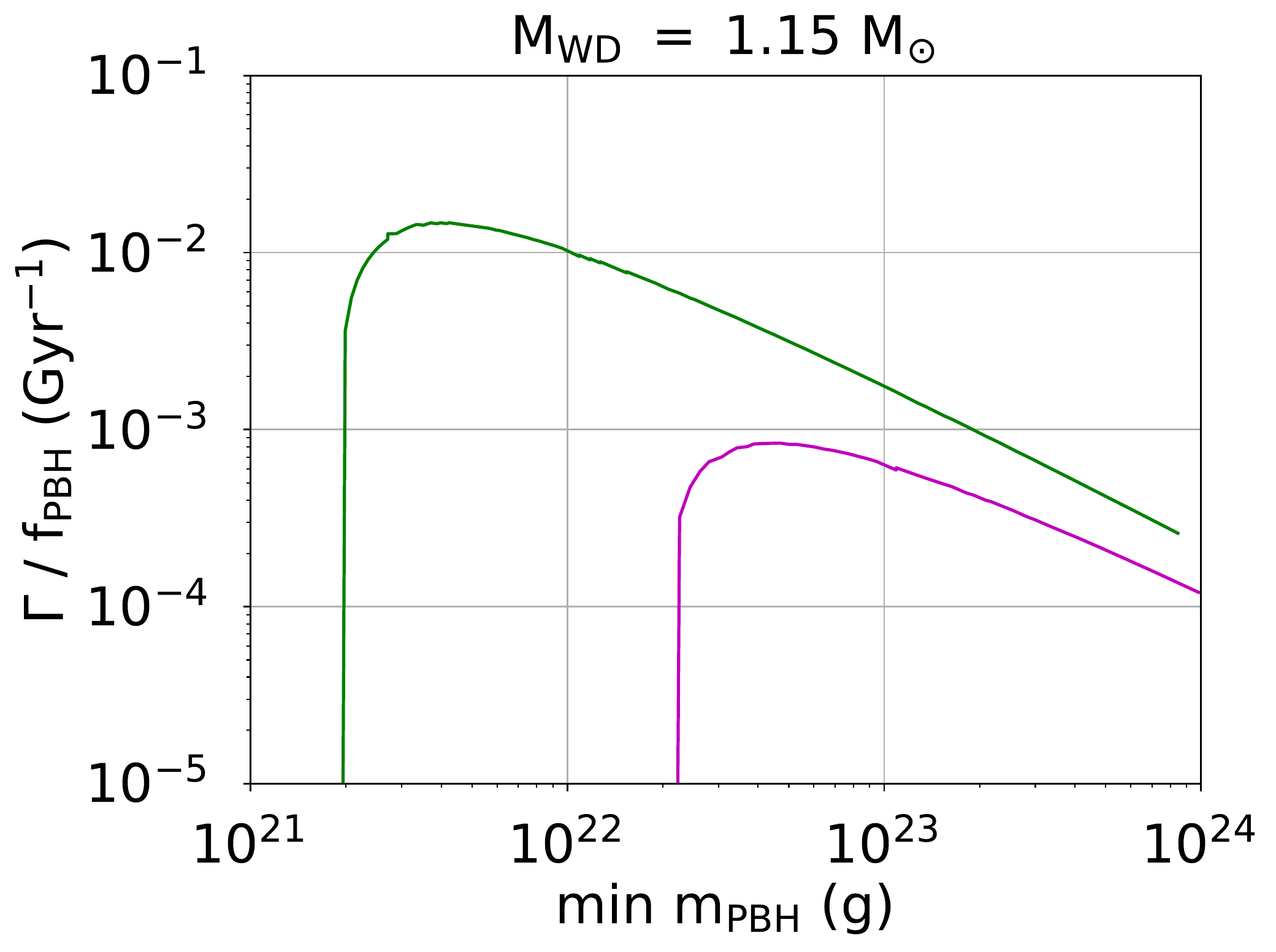}
\end{multicols}

\begin{multicols}{2}
    \centering
    \includegraphics[width=\columnwidth]{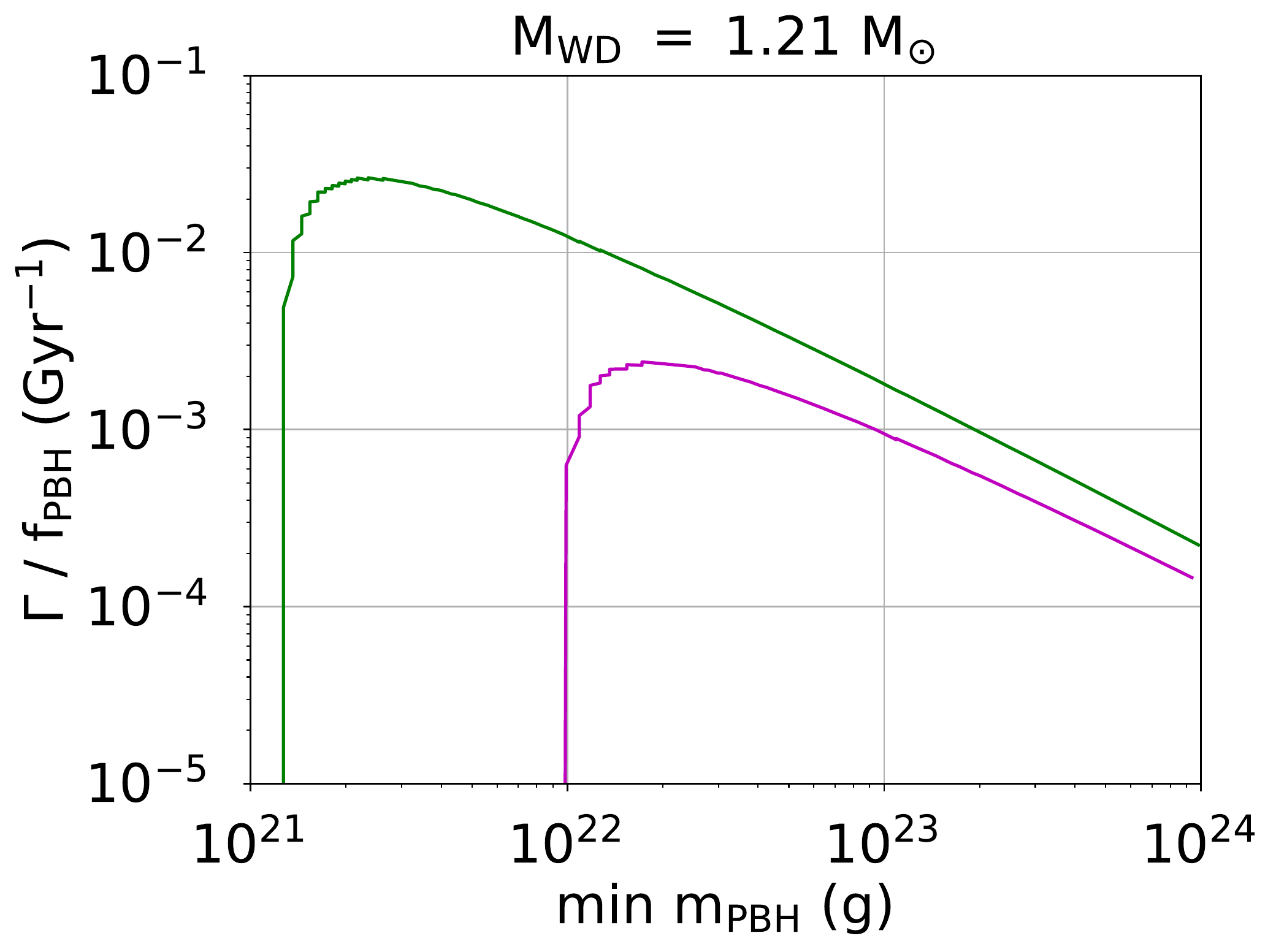}   
    \includegraphics[width=\columnwidth]{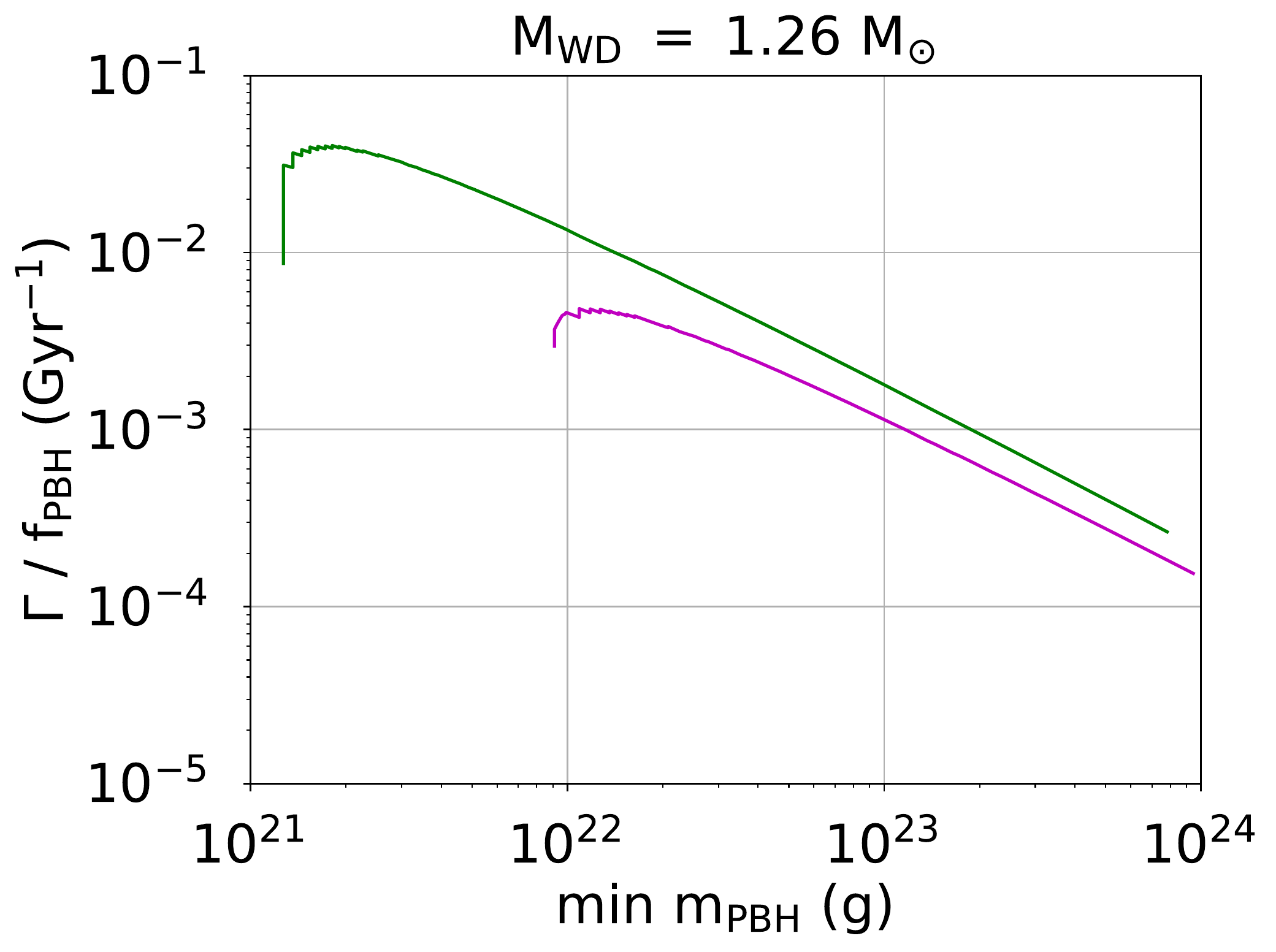}
\end{multicols}

\caption{The relationship between the rate of collisions $\Gamma$ and the minimum PBH mass $\mpbh$ to cause ignition ranging from $M_{\rm WD} = 1.00 - 1.26 M_{\odot}$. The upper (green) line is when we ignore the Kelvin-Helmholtz instability time as a criteria for ignition while the lower (purple) line is the rate when we include ``burning before instability'' as a criteria for ignition. These plots can also be interpreted as the PBH-induced ignition rates as a function of $\mpbh$.}
\label{fig:rate_plots}
\end{figure}

To constrain the fraction of PBHs that could be dark matter, we first need to calculate the rate of collisions between a WD and a PBH. The rate, most generally, is defined as
\begin{equation}
    \Gamma = \fpbh\frac{\rho_{\rm DM}}{m_{\rm PBH}}\int  P({\bf v}_{\infty})\sigma(v_{\infty}) v_{\infty}\, d^{3}{\bf v}_{\infty}
    = \pi \fpbh\frac{\rho_{\rm DM}}{m_{\rm PBH}} [\tilde l(r_{\rm min})]^2 \int  \frac{P({\bf v}_{\infty})}{ v_{\infty}}\, d^{3}{\bf v}_{\infty}.
\end{equation}
where $v_{\infty}$ is the velocity of the PBH at "infinity", $P(v_{\infty})$ is the probability, and $\sigma(v)$ is our cross-section. In the last equality, we assume the initial PBH velocity is small compared to the escape velocity from the surface of the WD ($5260 \ \rm{km/s}$ for $M_{\rm WD} =0.75 \ M_\odot$), so that $\sigma = \pi b_{\rm max}^{2}$ where $b \approx \tilde{l}(r_{\rm min})/v_{\infty}$ and $\tilde l(r_{\rm min})$ is the maximum specific angular momentum that leads to ignition.

We assume an offset Maxwellian velocity distribution, $P({\bf v}) = (2\pi\sigma_{\rm{rms}}^{2})^{-3/2} e^{-{({\bf v}-\bar{\bf v})^{2} / \sigma_{\rm rms}^{2}}}$; this gives
\begin{equation}
\int  \frac{P({\bf v}_{\infty})}{ v_{\infty}}\, d^{3}{\bf v}_{\infty}
= \frac1{\bar v} {\rm erf} \,\frac{\bar v}{\sqrt2\,\sigma}
\rightarrow \left\{ \begin{array}{lll} \sqrt{\frac 2\pi}\,\sigma^{-1} & & \bar v\ll\sigma \\ \bar v^{-1} & & \bar v\gg\sigma \end{array} \right.
\,,
\end{equation}
where erf is the error function.\footnote{We use the conventional normalization that ${\rm erf}\, \infty=1$. This integral can be easily performed by putting $\bar{\bf v}$ on the $z$-axis, and turning the integral in ${\bf v}$ into spherical polar coordinates. The $\phi$ and $\theta$ integrals are then trivial. The $v$ integral is then a Gaussian-type integral with limits not at $\pm\infty$, which is the defining form for the error function.} Taking $\sigma_{\rm rms} = v_{\rm gal} / \sqrt{2}$ and $\bar v = v_{\rm gal}$ (so for a white dwarf orbiting in the Galactic disk), the end result is that
\begin{equation}
    \Gamma = 2.65 \fpbh \frac{\rho_{\rm DM}}{\mpbh} \frac{\tilde{l}^{2}}{v_{\rm gal}}.
\label{eq:col_rate}
\end{equation}
We use parameters relevant to the Galactic disk: $v_{\rm gal} = 225$ km/s and $\rho_{\rm DM}c^2 = 0.4$ GeV/cm$^3$ \cite{2014JPhG...41f3101R}.

We calculated the rate of collisions given in Eq.~(\ref{eq:col_rate}) for each mass shell for each white dwarf in our mass range. However, to demonstrate our results graphically, we plotted six different WD masses in Fig.~\ref{fig:rate_plots} ranging from 1.00--1.26 $M_{\odot}$ with incriments of about $\Delta M_{\rm WD} \approx 0.05 M_{\odot} $. Since ignition was always possible with larger PBH masses past a lower cutoff, we calculated $\Gamma / \fpbh$ using the minimum PBH mass that causes ignition. This would provide the largest possible rate for igniting that shell because $\Gamma \sim 1/\mpbh$. As long as the minimum PBH mass to ignite a shell is an increasing function of $r$, we can also interpret this plot as an rate for PBH-induced ignitions as a function of $\mpbh$. (This is because at a given $\mpbh$, the indicated shell is the outermost shell in the WD that can be ignited, and hence the rate of PBH-induced ignitions is the rate of collisions reaching that shell.) For the most massive white dwarfs, there is a turnover due to carbon dissociation where the minimum PBH mass actually increases at small $r$; this region is not of interest since shells farther out would have ignited anyway.

In all cases, we found $\Gamma / \fpbh \sim {\rm few}\times 10^{-3} \  \rm{Gyr^{-1}}$ for the lower masses (up to $1.1M_\odot$), rising up to $\sim {\rm few}\times 10^{-2}$ Gyr$^{-1}$ at $M_{\rm WD}=1.26M_\odot$, when we don't impose the $\tau_{\rm burn}<\tau_{\rm KH}$ criterion, and the cases where we do impose this criterion were at least an order of magnitude less. Thus even the massive WDs survive for longer than a Hubble time, and more typical white dwarfs at $M_{\rm WD}\lesssim M_\odot$ can survive for many Hubble times. In particlar, the survival of the $1.28\pm 0.05$ $M_\odot$ white dwarf used for the constraint in Ref.~\cite{2015PhRvD..92f3007G} -- RX J0648.04418, which lies in the Milky Way's disk -- does not exclude $\fpbh=1$ at any $\mpbh$.

%We selected these cases from our full parameter space in Fig.~(\ref{fig:phase_space}) because they are commonly found white dwarf masses and demonstrate both a spectrum where we consider convection and where we don't. In our calculations, we found no cases where ignition was possible for white dwarfs with $M_{WD} \ \leq \ 0.95 \ M_{\odot}$. However, ignition was always possible when we don't consider it. In all cases, $\Gamma / f_{PBH} \lesssim \ 10^{-2} \ \rm{Gyr^{-1}}$ for cases where we don't consider convection, and the cases where we do consider convection were at least an order of magnitude less. Since the rate of collisions is so small and if we consider the probability of finding a white dwarf at some given mass, it is highly improbable that we can constrain the fraction of dark matter, $f_{PBH}$, that PBHs make up in the mass window of $10^{19}-10^{24}$ g. Therefore, previous constraints placed on this mass range must be lifted and reconsidered.

%\gabe{G.V.: I don't like this sentence. What should I substitute it with?}\chris{C.H.: See suggested revision; old version commented.}

As noted in Ref.~\cite{2015PhRvD..92f3007G}, there is a potential constraint on PBHs based on the rate of Type Ia supernovae {\em if} the ignition of a WD by a PBH can be shown to lead to a visible explosion rather than some other outcome (e.g., collapse, or explosion with insufficient $^{56}$Ni to be visible), and {\em if} the right environment can be found with sufficient dark matter and a sufficient density of high-mass C/O white dwarfs. However, we note that our minimum PBH masses to trigger an explosion in the conservative case where we require burning before mixing by Kelvin-Helmholtz instability (Fig.~\ref{fig:min_pbh}) are $\approx 5\times 10^{-12}M_\odot$, and a potential WD constraint will not probe below this. This is the same as the lower limit of the HSC M31 microlensing constraint, and thus the latter limit cannot be extended by the WD method if the $\tau_{\rm KH}$ criterion for ignition turns out to be necessary. If future work shows that this criterion is not necessary, then there is the potential to shrink the window for asteroid-mass PBH dark matter by one order of magnitude.

\section{Conclusion}
\label{sec:con}

\begin{figure}
    \centering
    \includegraphics{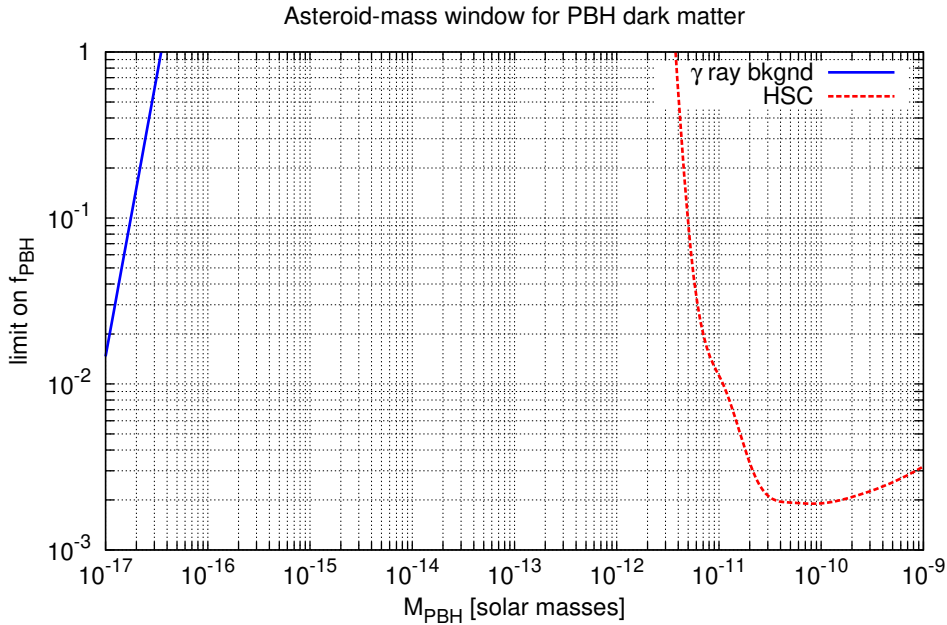}
    \caption{\label{fig:pbhnew}The asteroid-mass window for PBHs. The curves show upper limits from the $\gamma$-ray background \cite{2010PhRvD..81j4019C} and from HSC M31 microlensing \cite{2019NatAs.tmp..238N}. PBHs may make up all of the dark matter ($\fpbh=1$) for $3.5\times 10^{-17}M_\odot < \mpbh <4\times 10^{-12}M_\odot$.}
\end{figure}

In this paper, we have revisited constraints on asteroid-mass PBHs as a dark matter candidate.
%the sub-lunar-mass PBHs as dark matter candidates, i.e. $10^{-17}<m_{\rm PBH} < 10^{-8} M_{\odot}$, from the femtolensing of GRBs, the optical microlensing, the dynamical capture of PBH in stars, and the survival of WDs. 
We summarize our major work for each constraint below. 
\begin{itemize}
    \item \textit{Optical microlensing}. As one of the prime ways of constraining sub-solar-mass PBHs, the optical microlensing has been used to put constraints on PBHs down to $4\times 10^{-12}\,M_{\odot}$ using HSC observations of M31. We examined the scaling laws for the microlensing method in detail, and argue that due to a combination of finite source size and diffraction effects, it will be very difficult to extend this method to masses $\ll 10^{-12}M_\odot$.
    \item \textit{Dynamical capture}. Capture of PBHs in stars by dynamical friction and subsequent Bondi accretion may destroy the host stars. We use a phase-space argument to examine the rate of such captures, and find that survival of stars (including main sequence stars as well as neutron stars) does not significantly constrain PBHs. However, having {\em some} stellar disruptions, with a rate 3--4 orders of magnitude less than the supernova rate, is a robust prediction of asteroid-mass PBH dark matter. This presents a promising avenue for future study.
    \item \textit{White dwarf survival}. Another potential constraint on asteroid-mass PBHs is the possibility that in transiting a white dwarf, they may ignite the carbon and lead to destruction of the white dwarf, even if the PBH is not captured. We have simulated the thermal evolution of the materials along the trajectory before and after the shock passes, and examine whether the carbon ignition and the subsequent runaway explosion occur. We again conclude that at present there is not a constraint from white dwarf survival.
    \item Following these results and the re-analysis of the GRB femtolensing constraints \cite{2018JCAP...12..005K}, the window for low-mass PBHs to be {\em all} of the dark matter ($\fpbh=1$) extends from $3.5\times 10^{-17}M_\odot$ to $4\times 10^{-12}M_\odot$. This is bounded from below by $\gamma$-ray background constraints on Hawking radiation \cite{2010PhRvD..81j4019C} and from above by the revised HSC M31 microlensing constraints \cite{2019NatAs.tmp..238N}, as shown in Fig.~\ref{fig:pbhnew}.
\end{itemize}

There remain some significant astrophysical uncertainties in our calculations. In the case of dynamical capture of PBHs in stars, while the capture and subsequent Bondi accretion physics is simple, the behavior of stars in the final stages of their destruction is not. If sufficiently robust modeling of such a stage can be done and compared to observations, a constraint or detection of low-mass PBH dark matter would be possible. Moreover, this approach would cover the entire asteroid-mass window for PBH dark matter. In the case of white dwarf ignition, we extrapolated the fitting formulas for nuclear energy generation to higher temperatures; a more careful consideration of the rates, including a full reaction network for the higher temperatures, will be important in the future. However, we do note that the temperature rapidly falls after the shock passes, and most carbon burning occurs within the range of validity of the formulae. 
%\chris{Maybe clean this part up -- the discussion on WDs is a bit in the weeds for a conclusion.} \xiao{shortened} \chris{Do we like this version?} \paulo{I like it}\xiao{liked}

In conclusion, none of the mechanisms examined in this work can \textit{currently} exclude PBHs as dark matter candidates for the mass range $3.5\times 10^{-17}M_\odot$ to $4\times 10^{-12}M_\odot$.
%This is opposite to previous results ruling out this mass range due to femtolensing, WD survival, optical microlensing, and neutron star capture. 
However, we consider constraints that might arise from directly observable signatures of PBH-led stellar destruction as promising -- but requiring further studies in order to constrain this asteroid-mass window.  

%\chris{Someone should take a second pass through the conclusions; I think some of what I wrote is a bit redundant.} \paulo{[The last paragraph might be a little redundant, however, I am OK with it. I might try to make small changes on Friday.]}\xiao{the last paragraph serves as a veery brief summary, i think it's clear and useful.}

%Microlensing now lies as one of the prime ways of constraining PBH masses. This method of constraining $\mpbh$ is practical for masses $>10^{-11} M_{\odot}$ with current optical instruments. With future surveys, it will be possible to push down to masses $\sim 10^{-12} M_{\odot}$ (i.e WFIRST) \paulo{[Is this true?]}.

\appendix

\section{Some subtleties in the rate of capture of PBHs by stars}
\label{app:subtleties}

There are some potential caveats in the derivation of Eq.~(\ref{eq:NdotB}) that are worth noting, but that we do not expect to appreciably change our results.

The first is that $\ln\Lambda$ in Eq.~(\ref{eq:dfdtCh}) is not truly a constant: $\Lambda$ is a ratio of maximum to minimum impact parameter. The maximum depends on the global structure and oscillatory modes of the star, and the latter on the velocity of the PBH. This means that phase space density $f$ is not truly conserved. However, we expect this to be a small effect in practice. To see this, let us define $\tau_{\rm drag}=v/|\dot v_{\rm Ch}|$ to be the stopping time due to dynamical friction. We see that $\tau_{\rm drag} = v^3/(4\pi G^2\mpbh\rho\ln\Lambda)$. We may then re-write Eq.~(\ref{eq:dfdtCh}) as
\begin{equation}
\frac{{\rm d}f}{{\rm d}t} =  \frac1{\ln\Lambda}\frac{\partial \ln\Lambda}{\partial\ln v} \frac{f}{\tau_{\rm drag}}.
\end{equation}
Integrating gives
\begin{equation}
f = f_{\rm init} \exp \int \frac{1}{\ln\Lambda}
\frac{\partial \ln\Lambda}{\partial\ln v}
\frac{{\rm d}t}{\tau_{\rm drag}},
\end{equation}
instead of $f=f_{\rm init}$ (expected if phase space density is conserved). We expect $\partial\ln\Lambda/\partial\ln v$ to be of order unity, and if the PBH has not yet been stopped in the star, we expect $\int \tau_{\rm drag}^{-1}\,{\rm d}t\lesssim 1$. Thus the integral is expected to be at most of order $1/\ln\Lambda$, which would typically be $\sim 1/30$. Thus it should still be a good approximation to take $f\approx f_{\rm init}$.

Another concern is that -- while the dynamical friction formula gives a good description of the energy loss of a PBH passing through a star -- it fails spectacularly for a PBH on an orbit (bound or unbound) with periastron of a few stellar radii. In this case, the Chandrasekhar formula using the local matter density gives zero energy loss, but in fact the PBH can excite tides on the star (predominately through the $f$-modes) and lose energy. This mechanism of ``tidal capture'' \cite{1975MNRAS.172P..15F} has not been taken into account here.

We can investigate the significance of tidal capture by using the energy loss \cite{1977ApJ...213..183P}; the energy loss in a near-parabolic encounter at periastron distance $R_p$ due to tidal excitation of modes of multipole $\ell$ is
\begin{equation}
|\Delta E_\ell| = \frac{G\mpbh^2}{R} \left(\frac{R_p}{R}\right)^{-2\ell-2}T_\ell(\eta),
\label{eq:tidal}
\end{equation}
where $T_\ell$ is a dimensionless function that depends on the mode structure of the star and $\eta = (R_p/R)^{3/2}$ is a ratio of timescales; note that $\lim_{\eta\rightarrow\infty}T_\ell(\eta)=0$. Usually the quadrupole ($\ell=2$) and octupole ($\ell=3$) dominate, with higher $\ell$ contributing less.

While the tidal excitation picture is the formally correct way to calculate $\Delta E_\ell$, we can paste tidal capture into the dynamical friction formalism by introducing an ``effective'' density $\rho_{\rm eff}$ contributing a drag $F_{\rm drag} = 4\pi G^2\mpbh^2\rho_{\rm eff} (\ln\Lambda)/v^2$. We may choose this effective density to give the correct tidal energy loss, Eq.~(\ref{eq:tidal}), when the Chandrasekhar formula is used. If we hypothesize an effective density $\rho_{\rm eff} = A(3M/4\pi R^3)(r/R)^{-\alpha}$, then we can compute the energy loss $\int F_{\rm drag}\,ds$ along a parabolic trajectory. For a parabolic trajectory, we first note that the path length is ${\rm d}s = \sqrt{r/(r-R_p)}\,{\rm d}r$, and that the energy loss must be doubled to take into account the inward-going and outward-going arcs of the trajectory. Also in this case $v^2=2GM/r$. Then:
\begin{eqnarray}
|\Delta E_{\rm Ch,eff}| &=& 2\int_{R_p}^\infty 4\pi G^2 \mpbh^2 A \frac{3M}{4\pi R^3} \left( \frac{r}{R}\right)^{-\alpha} \frac{\ln\Lambda}{2GM/r} \, \sqrt{\frac{r}{r-R_p}}\,{\rm d}r
\nonumber \\
&=&  -\frac{3 G \mpbh^2 A R_p^2 }{ R^3} \left( \frac{R_p}{R}\right)^{-\alpha} \int_1^0 y^{\alpha-3} \ln\Lambda \, \sqrt{\frac{1}{1-y}}\,{\rm d}y
\nonumber \\
&=&  \frac{3 G \mpbh^2 A }{ R} \left( \frac{R_p}{R}\right)^{2-\alpha} B(\alpha-3,\tfrac12) \ln\Lambda,
\end{eqnarray}
where we have made the substitution $y=R_p/r$ and introduced the beta function $B$.

If $T_2(\eta)$ were constant at $\sim 0.25$ (see Fig.~1 of Ref.~\cite{1977ApJ...213..183P}, valid for an $n=3$ polytrope), then the above considerations for the quadrupole case would correspond to $\alpha = 8$ and $A\ln\Lambda = T_2/[3B(5,\tfrac12)] = 0.4T_2 \sim 0.1$. This overestimates the energy loss, especially at large $R_p$. However, even in this case, the effective density around the star has mass
\begin{equation}
M_{\rm eff} = \int 4\pi r^2\rho_{\rm eff}\,{\rm d}r
= \frac{3}{\alpha-3}AM =0.6AM \sim \frac{0.06M}{\ln\Lambda}.
\end{equation}
It is thus clear that the amount of effective mass that we need to introduce in order to ``mock up'' the tidal capture effect is tiny compared to amount of true mass in the star. Since we obtained a PBH capture rate that is proportional to the total mass, this means that tidal capture will lead only to a tiny increase in the rate at which the star captures PBHs. Therefore we ignore it in the main text.

\section{Equation of state}
\label{app:code_unit}

This appendix describes the equation of state of shock-heated white dwarf material, $P(\rho,\epsilon)$.

\subsection{Physical description}

We will consider the WD materials to be made of ultra-relativistic degenerate $e^-,e^+$, and (ideal gas) ions with average atomic number $\bar{Z}=7$ and mass $\bar{A}=14$ amu, which is roughly the case for C/O WD cores. We introduce a dimensionless parameter $J\equiv \mu_{e^-}/(k_B T)$ to describe the thermodynamics of the WD material (along with density $\rho$), where $\mu_{e^-}$ is the chemical potential for the electrons. For positrons, we have $\mu_{e^+} = -\mu_{e^-}$. The number densities of electrons and positrons are then given by
\begin{equation}
n_{e^\pm} = 2\int\frac{d^3p}{(2\pi\hbar)^3} \frac{1}{e^{[E(p)\pm\mu_{e^-}]/k_BT}+1} = - \frac{2}{\pi^2(\hbar c)^3}(k_BT)^3\,\li_3\left(-e^{\mp J}\right)~,
\end{equation}
where $\li_n(\cdots)$ is the $n$-th order polylogarithm function\footnote{The polylogarithm function is mathematically related to the complete Fermi-Dirac integral (see Eq.~5.4.1 in \cite{NIST:DLMF}), \ie
\begin{equation}
F_s(x)=\frac{1}{\Gamma(s+1)}\int_0^\infty \frac{t^s}{e^{t-x}+1}dt = -\li_{s+1}(-e^x)~,~~(s>-1)~.
\end{equation}
A useful limiting case is given in \cite{Wood110},
\begin{equation}
\lim_{\Re(x)\rightarrow\infty}\li_s(-e^x) = -\frac{x^s}{\Gamma(s+1)}~,~~(s\neq -1,-2,-3,\cdots)~,
\end{equation}
which tells us that positrons are negligible at low temperatures.}, and the second equality has used the fact that $E(p)=pc$ for ultra-relativistic particles. With the electrical neutrality, the number density of ions is then $n_{\rm ion} = (n_{e^-} - n_{e^+})/\bar{Z}$. Since the mass is dominated by the ions, we obtain the density of the material as
\begin{equation}
\rho = n_{\rm ion}m_{\rm nuc}\bar{A} = A_f\mu_{e^-}^3 f(J)~,
\label{eq:rho_in_fJ}
\end{equation}
where $m_{\rm nuc}= 1.67\times 10^{-27}\,$kg is the mass of nucleon, $A_f = m_{\rm nuc}\bar{A}/[\bar{Z}\pi^2(\hbar c)^3]$, and $f(J)=-2\left[\li_3\left(-e^{J}\right)-\li_3\left(-e^{-J}\right)\right]/J^3~$.
Similarly, we find the total pressure from electrons, positrons and ions is given by
\begin{equation}
P = A_g g(J)\mu_{e^-}^4 + n_{\rm ion}\frac{\mu_{e^-}}{J} = \mu_{e^-}^4\left[A_g g(J) + \frac{A_f}{m_{\rm nuc}\bar{A}}\frac{f(J)}{J}\right]~,
\end{equation}
where $A_g= 1/[\pi^2(\hbar c)^3]$, and $g(J)= -6/\left[\li_4\left(-e^{J}\right)+\li_4\left(-e^{-J}\right)\right]/J^4~$.
Denoting the total specific energy (internal + kinematic energy per mass) of electrons, positrons, ions, and their sum as $\epsilon_{e^-},\epsilon_{e^+},\epsilon_{\rm ion},\epsilon$, respectively, we can write $e^\pm$ total energy density as
\begin{equation}
\rho\epsilon_{e^\pm} = 2\int\frac{d^3p}{(2\pi\hbar)^3} \frac{E(p)}{e^{[E(p)\pm\mu_{e^-}]/k_BT}+1} = -\frac{6}{\pi^2(\hbar c)^3}(k_BT)^4\,\li_4\left(-e^{\mp J}\right)~.
\end{equation}
Thus, the total energy density of the material is given by
\begin{equation}
\rho\epsilon = \rho(\epsilon_{e^-}+\epsilon_{e^+}+\epsilon_{\rm ion}) = A_g g(J) \mu_{e^-}^4 + \frac{3}{2}\frac{\rho}{m_{\rm nuc}\bar{A}}\frac{\mu_{e^-}}{J}~,
\label{eq:erg_density}
\end{equation}
where the last term comes from the ideal gas law. Combining Eqs.~(\ref{eq:rho_in_fJ}) and (\ref{eq:erg_density}), we find
\begin{equation}
\epsilon\rho^{-1/3} = A_h h(J)~,
\label{eq:hJ}
\end{equation}
with the coefficient $A_h$ and the dimensionless function $h(J)$ defined as
\begin{equation}
A_h \equiv \frac{3\hbar c\bar{Z}^{1/3}}{(m_{\rm nuc}\bar{A})^{4/3}} \left(\frac{\pi^2}{2}\right)^{1/3}~,~~h(J)\equiv \left[\frac{\bar{Z}}{3}\frac{g(J)}{f(J)}+\frac{1}{2J}\right]\left[\frac{2}{f(J)}\right]^{1/3}~.
\end{equation}
This set of equations (Eqs.~\ref{eq:rho_in_fJ}--\ref{eq:hJ}) suggests that the thermodynamic state of the material can be fully determined by only two variables, $\rho$ and $J$. Since $h(J)$ is a monotonically decreasing function, given $\rho$ and $\epsilon$, one can determine $J$ from $\rho$ and $\epsilon$ by inverting Eq.~(\ref{eq:hJ}) numerically. Then, the electron chemical potential can be calculated by inverting Eq.~(\ref{eq:rho_in_fJ}), \ie, $\mu_{e^-} = \lbrace\rho/[A_f f(J)]\rbrace^{1/3}~$, and from there, we immediately have access to the pressure, temperature, and other thermodynamic quantities.

\subsection{Conversion to code units}

In the simulation, we adopt the unit system $c_{s,\infty}=\rho_\infty = R_c = 1$. In these units, we have initial pressure $P_{\rm init}=1/\gamma = 3/4$, and specific internal energy $\epsilon_{\rm in,init} = 1/[\gamma(\gamma-1)]=9/4$.

In order to express all the quantities in the code units, we need to express our basic units $\rho_\infty,c_{s,\infty}$ in SI units. Defining a dimensionless Fermi momentum $x$ as $x\equiv p_{\scriptscriptstyle \rm F}/(m_e c)$, where $p_{\scriptscriptstyle \rm F}$ is the Fermi momentum and $m_e$ is the rest mass of electron, we have
\begin{align}
\rho_{\infty} =& \frac{\bar{A}m_{\rm nuc}}{3\pi^2\bar{Z}} \left(\frac{m_e cx}{\hbar}\right)^3 \equiv 1~,~~P_{\infty} = \frac{\hbar c}{12\pi^2}\left(\frac{m_e cx}{\hbar}\right)^4~,~~c_{s,\infty}^2 = \frac{4P_{\infty}}{3\rho_{\infty}} = \frac{\bar{Z}\hbar c}{3\bar{A}m_{\rm nuc}} \frac{m_e cx}{\hbar} \equiv 1~.
\end{align}
By eliminating $x$, $A_h$ is reduced to $A_h = 9/(6^{1/3}\bar{Z})$ in the code units. To further simplify the problem, we rescale the electron chemical potential $\mu_{e^-}$ by the background electron Fermi energy $E_{{\scriptscriptstyle\rm F},\infty}$, which is also equal to the background electron chemical potential, \ie
\begin{equation}
E_{{\scriptscriptstyle\rm F},\infty} = \mu_{e^-,\infty} = \left[\frac{\rho_\infty}{A_f f(\infty)}\right]^{1/3}~.
\end{equation}
Note that when $J\rightarrow\infty$, $f(\infty)=1/3$, $g(\infty)=1/4$, so we have
\begin{equation}
\frac{\mu_{e^-}}{E_{{\scriptscriptstyle\rm F},\infty}} = \left[\frac{\rho/\rho_\infty}{3f(J)}\right]^{1/3} \xrightarrow{\rm code}\left[\frac{\rho}{3f(J)}\right]^{1/3}~.
\end{equation}
Similarly, the pressure is
\begin{equation}
P=\frac{P}{P_{\infty}}\frac{3\rho_\infty c_{s,\infty}}{4}=3\rho_\infty c_{s,\infty}\left[g(J)+\frac{f(J)}{\bar{Z} J}\right]\left(\frac{\mu_{e^-}}{E_{{\scriptscriptstyle\rm F},\infty}}\right)^4 \xrightarrow{\rm code}3\left[g(J)+\frac{f(J)}{\bar{Z} J}\right]\left[\frac{\rho}{3f(J)}\right]^{4/3}~.
\end{equation}
Thus, coefficients $A_f$ and $A_g$ are eliminated from the problem.

\acknowledgments
PMC and CMH are supported by the Simons Foundation, the US Department of Energy, the NSF, and NASA. XF is supported by NASA ROSES ATP 16-ATP16-0084 and NASA ADAP 16-ADAP16-0116 grant. 
We thank Yacine Ali-Ha\"imoud, John Beacom, Matthew Penny, Annika Peter, Tuguldur Sukhbold, and Masahiro Takada for useful feedback and suggestions on models used in this paper. We thank Masahiro Takada for providing the numerical data for the HSC constraint shown in Fig.~\ref{fig:pbhnew}. We also thank an anonymous referee for useful comments which improved the paper.

%PMC and XF are grateful to Tuguldur Sukhbold and John Beacom for useful discussions. GV is thankful to Annika Peter for useful suggestions. CMH thanks Yacine Ali-Ha\"imoud and Masahiro Takada for useful conversations. 

\bibliographystyle{JHEP.bst}
\bibliography{pbh.bib}

\end{document}